\documentclass{elsart}
\usepackage{epsfig}

\begin{document}

\begin{frontmatter}

\title{ Isospin Effects in Nuclear Fragmentation}

\author
{ V.Baran${}^{{1,2}}$, M.Colonna${}^{{1}}$, 
M.Di Toro${}^{{1}}$, V.Greco${}^{{1}}$}
\author{M.Zielinska-Pfab\'e${}^{{3}}$
 and H.H.Wolter${}^{{4}}$
}
\address{1) Laboratori Nazionali del Sud, Via S. Sofia 44,
I-95123 Catania, Italy}
\address{and University of Catania}

\address{2) NIPNE-HH, Bucharest, Romania}

\address{3) Smith College, Northampton, USA}

\address{4) Sektion Physik, University of Munich, Germany}


\begin{abstract}
We investigate properties of the symmetry term 
in the equation-of-state (EOS) of nuclear matter (NM) 
from the analysis of simulations of fragmentation events 
in intermediate energy heavy ion collisions. 
For charge asymmetric systems a qualitative new feature
in the liquid-gas phase transition is predicted: the onset 
of chemical instabilities with a 
mixture of isoscalar and isovector components.
This leads to a separation into a higher density (``liquid'') 
symmetric and a low density (``gas'') neutron-rich phase, 
the so-called  neutron distillation effect.
We analyse the simulations with respect to the time evolution 
of the isospin dynamics, as well as with respect to the 
distribution and asymmetry of the final primary fragments.
Qualitatively different effects arise in central collisions, with
bulk fragmentation, and peripheral collisions with neck-fragmentation.
The neck fragments produced in this 
type of process appear systematically more neutron-rich from a dynamical
nucleon migration effect which is very sensitive to the symmetry
term in regions just below normal density.

In general the isospin dynamics plays an important role in all
the steps of the reaction, from prompt nucleon emission to
the sequential decay of the primary fragments. A fully microscopic
description of the reaction dynamics including stochastic
elements to treat fluctuations realistically is absolutely
necessary in order to extract precise information on the
fragmentation and the nuclear equation of state.
We have performed simulations for fragment production events in 
$n$-rich ($^{124}Sn$) and $n$-poor ($^{112}Sn$) symmetric colliding systems. 
We test the dependence of the isospin dynamics on the isospin EOS and
on the neutron enrichment of the system. 
\end{abstract}

\begin{keyword}
Asymmetric nuclear matter \sep Isospin dynamics \sep Fragmentation

\PACS 21.65.+f \sep 25.70.Pq \sep 25.70.Mn \sep 24.10.Cn
\end{keyword}
\end{frontmatter}

\section{Introduction}
The availability of exotic (radioactive) beams has driven 
a strong interest in nuclear structure studies of $\beta-$unstable
nuclei. It is clear that essential complementary information
will come from charge asymmetry effects on non-equilibrium 
nuclear dynamics of heavy ion collisions. A quantitative analysis of 
this aspect, trying to
identify  some  sensitive observables in the 
reaction mechanisms, is the aim of the present work.

We will show that it should be possible to extract
 information on the symmetry term
of the nuclear EOS in regions away from normal density
under laboratory controlled conditions.
Asymmetric nuclear matter models at high density 
have been tested so far only in astrophysics
contexts, in particular in supernovae explosions and neutron stars
\cite{irv78,lat91,ptt92,sumi94,pet95,lee96}.
Although in heavy ion collisions at intermediate
energies we certainly cannot reach very high density regions,
we should be able to obtain information on the slope
of the symmetry term, i.e. the {\it Symmetry Pressure},
below the saturation density.
We then can put experimental constraints on
the effective interactions used in astrophysical contexts 
\cite{bom94,pra97}.
Moreover we like to remark that the same symmetry pressure
is of relevant importance for structure properties, being
clearly linked to the thickness of the neutron skin in
n-rich (stable and/or unstable) nuclei (see \cite{pr95} and the discussion
in refs. \cite{bro00,hor01}).

There are very stimulating predictions on new 
phases of asymmetric nuclear matter  that eventually could be reached in
heavy ion reaction dynamics with radioactive beams.
 The onset of coupled mechanical and chemical 
instabilities is envisaged \cite{mue95,bao197,cdl98,bar01}, that should 
lead to interesting
experimental signatures of the dynamics
of this new phase transition. In particular, compared to statistical 
models  more symmetric and narrower isotopic
distributions of primary fragments are predicted, which should be 
sensitive to the symmetry term of the EOS.
For semi-peripheral collisions the isospin dynamics of the participant
zone is seen to be affected by the symmetry term \cite{fab98}
with respect to 
neck-dynamics, mid-rapidity fragmentation and
fast-fission of the spectators \cite{neck,tok96,dur98,indra,col00}. 
We will see that independent new information on the symmetry 
term can be derived.

In the last years some first data have appeared on
isospin effects in reaction dynamics with a few theoretical analyses
(see  the recent reviews 
\cite{wes98,bkb98,eri98,sob99}).
Although the data are mostly of inclusive type and the theoretical
studies have not focussed on the effect of different symmetry
terms, a noticeable dependence of 
the reaction mechanism on the charge asymmetry emerges clearly.
Very recently accurate results from high-performance 
$4\pi$-detectors have appeared \cite{xu00,vercr,ts01,yencr,schcr,chimera,desouza}
that are strongly stimulating theoretical interpretations \cite{bao00,ditcr}. 
These are the main motivations of this work.

In section 2 we discuss the wide range of predictions that still
exist on the density
dependence of the symmetry term in nuclear matter. Section 3 reviews the 
results on new features of the
liquid-gas phase transition in asymmetric NM  and
the isospin distillation effect. In sect. 4 the isospin dependent transport
equations for the collision dynamics are introduced
with special attention to the construction of the stochastic
term. Section 5 is devoted to a detailed discussion of the "ab initio" 
collision simulations for $n$-rich and $n$-poor systems using different 
equations of state.
 A summary of the main results with related
perspectives is given in Section 6.

\section{Symmetry term effects on compressibility, saturation density
and the nucleon mean field}

A key question in the physics of
unstable nuclei is the knowledge of the EOS for asymmetric nuclear
matter away from normal conditions. 
We recall again that the symmetry term at low densities
affects the neutron skin structure, while in the 
high density region it is crucial for supernovae dynamics
and neutron star cooling. The paradoxial situation is that while we
are planning second and third generation facilities for
radioactive beams our basic knowledge of the symmetry
term of the EOS is still extremely poor.
Effective interactions are obviously tuned to symmetry properties
around normal conditions and extrapolations are very uncertain. 
Microscopic approaches based on realistic NN
interactions, Brueckner or variational schemes, or on effective
field theories show a large range of predictions.
As an example, in fig.1 we show  the isospin dependence of some EOS's
which, however,  have the identical saturation properties for
symmetric NM: $SKM^*$ \cite{kri80}, $SLy230b~(SLy4)$ 
\cite{lyon97} and
$BPAL32$ \cite{bom94,rho}.

\begin{figure}[htp]
\epsfysize=6.cm
\vspace{5cm}
\centerline{\epsfbox{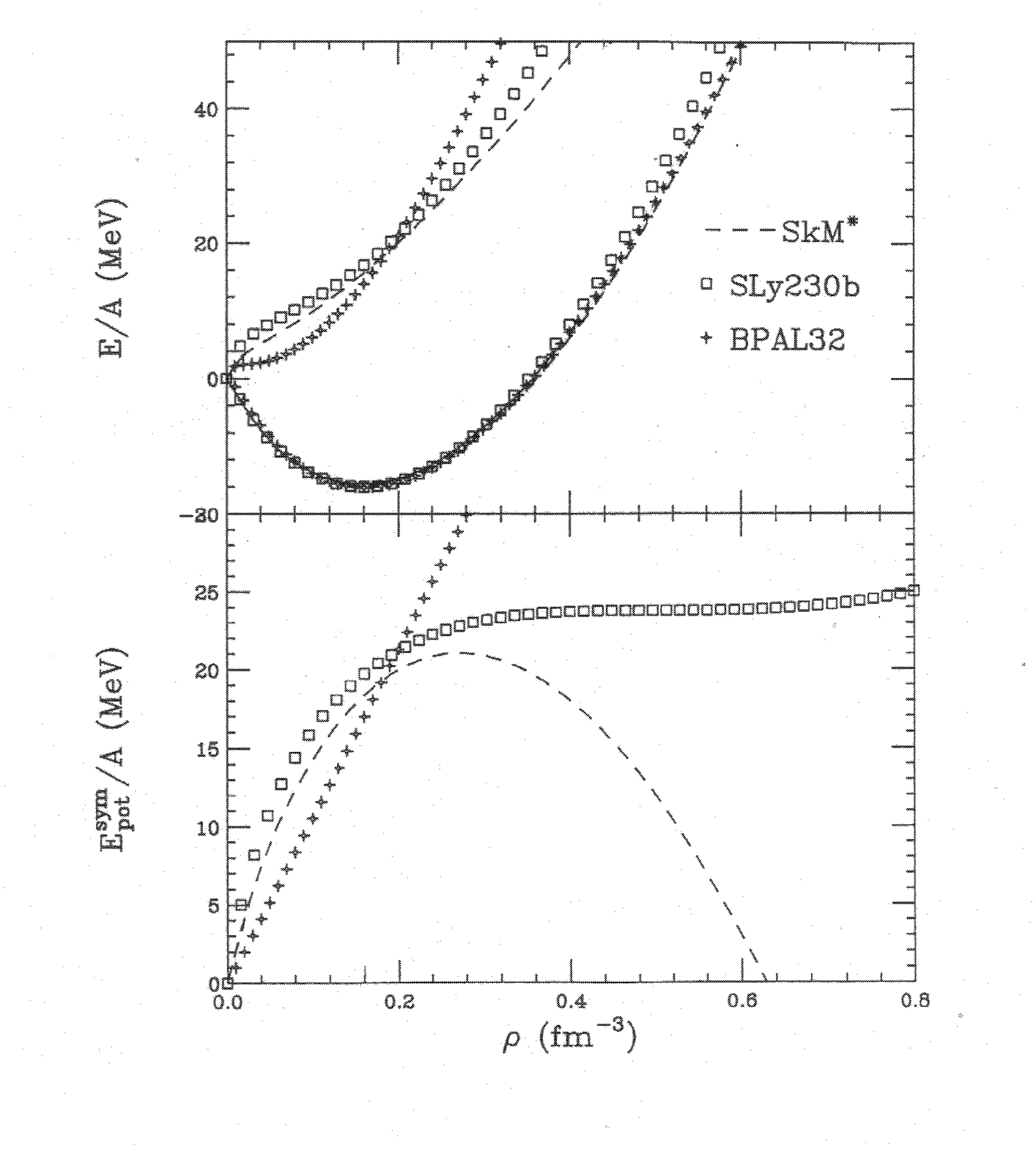}}
\caption{\it EOS
for various effective forces.
Top panel: neutron matter (upper curves), symmetric matter
(lower curves); Bottom panel: symmetry term of potential energy.}
\end{figure}

In the following we
will refer to an "asy-stiff" EOS when we are 
considering a potential symmetry term linearly increasing with nuclear 
density and to a "asy-soft" EOS when the symmetry term shows a 
saturation and sometimes even a decrease at higher densities \cite{eri98}.
In some cases, in order to enhance the dynamical effects, we will
consider also an "asy-superstiff" behaviour with a roughly
parabolic increase of the
 symmetry term with  density
\cite{pra97,bao00,rho}.
In the nuclear equation of state the symmetry term 
is written in the form
\begin{equation}
{E \over A} (\rho,I) = {E \over A} (\rho) + {E_{sym} \over A} (\rho)
 ~ I^2
\end{equation}
with $I=(N-Z)/A$ the asymmetry parameter. The  kinetic and potential 
contributions are
\begin{equation}
\epsilon_{sym} \equiv {E_{sym} \over A} (\rho) = {\epsilon_F(\rho) \over 3} 
+ {C(\rho) \over 2} ~ {\rho \over \rho_0}
\end{equation}
In a Skyrme-like parametrization of effective interactions the function
$C(\rho)$ has the form:
\begin{eqnarray}
{C(\rho) \over {\rho_0}} = - {1 \over 4} \Big[t_0 (1 + 2 x_0) + {t_3 \over 6} 
(1 + 2 x_3)~ \rho^\alpha \Big] + \nonumber\\
{1 \over {12}}
\Big[t_2 (4 + 5 x_2) - 3 t_1 x_1 \Big] \Big( {{3 \pi^2} \over 2} \Big)
^{2/3}~ \rho^{2/3}
\end{eqnarray}
We note that the second term is related to isospin effects in the momentum 
dependence \cite{lyon97}.

In fig.1 (bottom) we report the density dependence of the potential symmetry 
contribution, i.e. the second term of eq.(2), for the three different 
effective interactions shown in the upper panel.
While all curves obviously cross at normal density $\rho_0$, quite 
large differences are seen with repect to  values and slopes
in low density and particularly in high density
regions. We note especially the uncertainty on the symmetry pressure
even around $\rho_0$, which is of great importance for structure calculations,
as mentioned in the introduction.

We start to discuss some simple considerations
of asymmetry effects in infinite matter on compressibility and saturation
density, observables related to monopole resonances
and to the bulk density in a heavy nucleus \cite{ysk98}.  
From a linear expansion of the energy around the symmetry value we 
obtain for
the variation of saturation density with asymmetry
\begin{equation}
\Delta \rho_0(I) = - {{9\rho_0^2} \over K_{NM}(I=0)}~
{d \over {d\rho}} {E_{sym} \over A} (\rho) \Big\vert
_{\rho=\rho_0} I^2 ~~~<~0\quad,
\end{equation}
where $K_{NM}(I=0)$ and $\rho_0$ are respectively compressibility
and saturation density of symmetric NM.
Eq.(4) has an intuitive {\it geometrical} interpretation.
Asymmetry results in an extra pressure 
$P_{sym}= {\rho}^2 d\epsilon_{sym}/d\rho$
that can be compensated by moving to the left the saturation
point ($P=0$) by the quantity  $\Delta\rho_0$.
For the compressibility shift we have, after some algebra,
\begin{equation}
\Delta K_{NM}(I) = 9\rho_0 \Big[ \rho_0 {d^2 \over {d\rho^2}}
~-~2 {d \over {d\rho}} \Big] {E_{sym} \over A} (\rho) \Big\vert
_{\rho=\rho_0} I^2 ~~~<~0\quad,
\end{equation}
to note the interplay between slope and curvature of the symmetry term.

 The predictions are quite different even
for relatively small asymmetries. E.g., as expected from fig.1,
 $SKM^*$ gives the 
largest variation for the compressibility and the smallest for the saturation
density \cite{ysk98}.
Thus there are good chances of a direct experimental observation. 

From eqs. (1-3) we can derive a general Skyrme-like form for 
neutron and proton mean field 
potentials
\begin{equation}
U_q =
  A\left({\rho_s \over \rho_0}\right) 
+ B\left({\rho_s \over \rho_0}\right)^{\alpha +1} 
+ C\left({\rho_i \over \rho_0}\right)\tau_q
+ {1 \over 2} {\partial C \over \partial \rho_s}
{\rho_i^2 \over \rho_0}~,
\end{equation}
where $\rho_s \equiv \rho_n + \rho_p$ and $\rho_i \equiv \rho_n - \rho_p$
are respectively isoscalar and isovector densities, 
and $\tau_q$ = +1 ($q=n$), respectively  $\tau_q$ = -1 ($q=p$).
The symmetry contribution to the mean field (second part of eq.(6)) 
for the  parametrizations discussed 
in this work has the following form:
For the "asy-soft" EOS we use
the $SKM*$ parametrization shown
in fig. 1. For the "asy-stiff" EOS the density dependence is given
by $C(\rho) = const \simeq 32 MeV$. For the "asy-superstiff" EOS 
we use a symmetry term rapidly increasing around normal density
with a form $2 {\rho}^2 /\rho_0(\rho + \rho_0)$ \cite{bao00,rho}.
The  density dependence of the neutron and proton mean fields 
for these parametrizations are shown in Fig.2 
for a system with $N=1.5~ Z$ ($I=0.2$). 

\begin{figure}[htp]
\epsfysize=5.cm
\vspace{6cm}
\centerline{\epsfbox{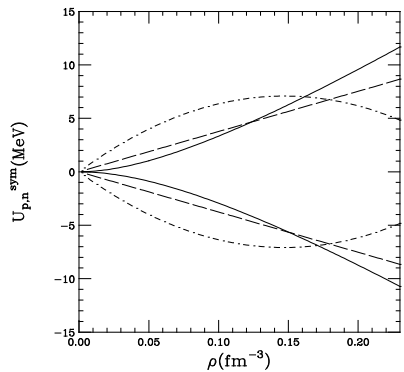}}
\caption{\it Symmetry contribution to the mean field at $I=0.2$
for neutrons (upper curves) and protons (lower curves):
dashed lines "asy-soft", solid
lines "asy-stiff", long dashed lines "asy-superstiff"}.
\end{figure}

In regions just off normal
density the field "seen" by neutrons and protons in the three
cases is very different, especially below normal density. 
We thus expect important transport effects during the reaction
at intermediate energies since the interacting asymmetric nuclear matter
will experience compressed and expanding phases before forming
fragments around normal density conditions.

A complementary and maybe more complete picture of
the isospin dynamics can be obtained from the analysis
of the density dependence  of the
neutron/proton chemical potentials 
$\mu_q \equiv \partial \epsilon (\rho_q, \rho_{q'}) / \partial \rho_q$,
$\epsilon$ being the energy density. 
We recall that the chemical
potentials contain all the contributions to the energy per
particle, including the scalar part and the kinetic symmetry
term in asymmetric matter. In this sense their study is
quite natural for the energetic arguments, while the symmetry mean field
of fig.2, although very instructive, shows only the potential part.
We note that in  non-equilibrium processes the mass flow
is determined by the differences in the local values of chemical
potential and it is directed from the regions of higher chemical potential to
regions of lower values until equalization. This is analogous to what
happens with the heat in a temperature gradient.
For a two component systems such discussion should be performed for 
each species.

\begin{figure}[htp]
\epsfysize=5.cm
\vspace{5cm}
\centerline{\epsfbox{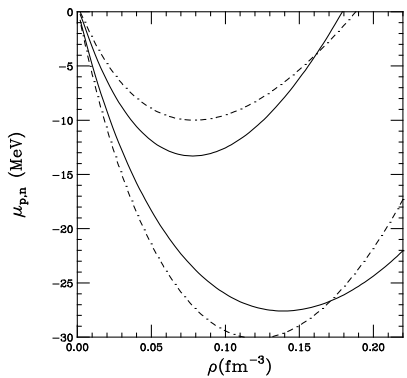}}
\caption{\it Density dependence of proton (lower curves) and neutron 
(upper curves) chemical potentials
for an asy-superstiff (solid lines) and asy-soft (dashed line) EOS for
asymmetry I=0.2.}
\end{figure}

In fig. 3 we show the proton and neutron chemical potentials for symmetric
and asymmetric ($I=0.2$) nuclear matter for two isospin EOS (soft, superstiff). 
We note that the difference between neutron and proton chemical potentials
is given by the relation
\begin{equation}
\mu_n - \mu_p = 4 \frac{E_{sym}}{A} I .
\end{equation}
It thus reflects directly the density dependence of the symmetry
potential as seen in the lower panel of Fig. 1. Two regions with 
significantly different behaviour can be distinguished below normal
density. In the region $\rho \leq .08 fm^{-3}$ neutrons and protons will
move both to higher densities, while in the region above and up to 
density $\rho_0$
neutrons will move to lower and proton to higher densities. This will
be seen to be useful for the interpretation of the reaction dynamics.

\section{Mechanical and Chemical Instabilities in Dilute 
Nuclear Matter: Isospin Distillation}

For charge asymmetric systems we expect qualitatively new features
in the liquid-gas phase transition; the onset of chemical instabilities
that will show up in a novel nature of the unstable modes, given by a 
mixture of isoscalar and isovector components.
In this section we will review the results in nuclear matter as given in
refs. \cite{mue95,bar01}.
In the framework of Landau theory for two component Fermi liquids the
spinodal border is determined by studying the stability of collective modes 
described by two coupled Landau-Vlasov equations for protons and neutrons.
In terms of the appropriate Landau parameters the stability condition can be 
expressed as \cite{bar98}
\begin{equation}
(1 + F_0^{nn})(1 + F_0^{pp}) - F_0^{np}F_0^{pn} > 0~.  
\label{eq:Land}
\end{equation}  
It was shown in \cite{bar01} that this condition is equivalent to 
the following thermodynamical condition
\begin{equation}
\left({\partial P \over \partial \rho}\right)_{T,y}
\left({\partial\mu_p \over \partial y}\right)_{T,P} > 0
\label{eq:ser}                                                   
\end{equation}
discussed in \cite{mue95,bar01,landau}, where $y$ is the proton fraction
and $\mu_p$ the proton chemical potential.       
In fig.4 we show the spinodal boundaries obtained from eq.~(\ref{eq:Land})
(continuous line with circles).
 The calculations are performed with the non-relativistic
Skyrme-like force $SKM^*$ \cite{bar01}, but very similar results can be
obtained with relativistic mean field approaches \cite{mue95}.
For asymmetric nuclear matter these boundaries are 
seen to contain the region  corresponding to
"mechanical instability",
$\left({\partial P \over \partial \rho}\right)_{T,y}<0$
(crosses). Outside, in the area between crosses and circles, the 
instability of the system is thus driven by 
the {\it chemical} condition 
$\left({\partial \mu_p \over \partial y}\right)<0$. 

We note, however, that just from  the above stability 
conditions
we cannot determine the nature of the fluctuations against which a binary
system becomes chemically unstable. Indeed, the thermodynamical
condition in eq.~(\ref{eq:ser}) cannot distinguish between two  very different 
situations which can be encountered in nature: an attractive
interaction between the two components of the mixture 
($F_0^{np},F_0^{pn} < 0$),
 as is the case of nuclear matter, or a  repulsive interaction
between the two species.
We define  density fluctuations as isoscalar-like in
the case when proton and neutron Fermi spheres
(or equivalently the proton and neutron densities) fluctuate in phase and as
isovector-like when the two Fermi spheres fluctuate
out of phase. Then it was shown, based on 
a thermodynamical approach to asymmetric Fermi liquid mixtures \cite{bar01},
that
chemical instabilities are triggered by isoscalar-like fluctuations 
in the first, 
i.e. attractive, situation
and by isovector-like fluctuations in the second one. 
For the dilute asymmetric nuclear
matter case, because of the attractive interaction between protons and neutrons 
at low density,
the phase transition is thus due to isoscalar-like fluctuations that induce
chemical instabilities while the system is never unstable against isovector
fluctuations. 

The apparent paradox that chemical instabilities are due to isoscalar-like
fluctuations can be understood from the behaviour of the chemical 
potenials in fig. 3. In the low-density region, where the instability 
occurs, both proton and neutron chemical potentials decrease
with density thus leading to an isoscalar-like fluctuation. However,
the slope of the chemical potentials is different for protons and 
neutrons, thus leading to a more proton-rich denser phase ("liquid")
and a ``chemical'' effect, the neutron distillation.   

Of course the same attractive interaction is also at the origin of phase
transitions in symmetric nuclear matter. However, in the asymmetric case
isoscalar fluctuations lead to a more symmetric high density phase
everywhere under the instability line defined by 
the eqs. (\ref{eq:Land},\ref{eq:ser})~ \cite{bar01,bar98}
and therefore to a more neutron-rich gas ({isospin distillation}).
The mechanical instability zone
shown in figs.3b,3c for asymmetric cases, has not a real physical
meaning: the same isospin distillation effect also happens
if the system  is prepared inside the
mechanical instability region. Thus there is a smooth transition
from "chemical" to "mechanical" instability \cite{bar01}.

\begin{figure}[htp]
\epsfysize=5.cm
\vspace{5cm}
\centerline{\epsfbox{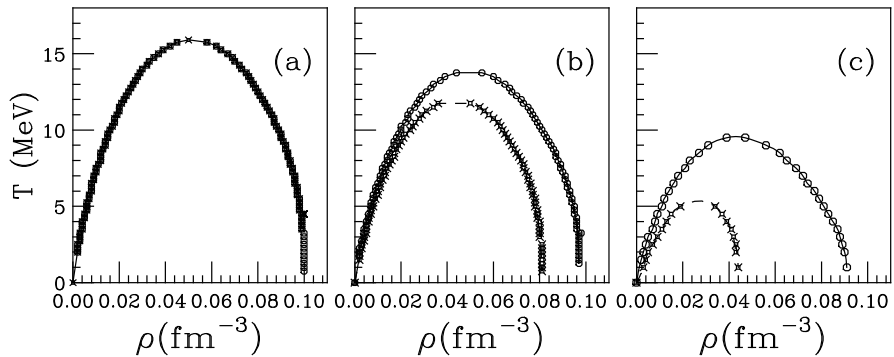}}
\caption{\it Spinodal boundary of asymmetric nuclear
matter (open circles) and mechanical instability boundary (crosses) for three
proton fractions: a) $y=0.5$, b) $y=0.25$, c) $y=0.1$}
\end{figure}

Since during the dynamics of a collision
the system can deeply enter the instability region it is important
to have a more detailed information on the space-time structure
of the unstable modes leading to fragment formation. This can be
discussed in the Landau kinetic approach to the
linear response theory \cite{landau,bay91}. The Landau dispersion relations
have indeed imaginary sound velocity solutions with well defined
structure in the ($\delta\rho_n, \delta\rho_p$) space
\cite{cdl98,bar98}.
They always lead to the prediction of a
very neutron rich gas phase versus a more symmetric liquid phase formed 
in a dynamical non-equilibrium mechanism on short time scales.
This "chemical effect" is very sensitive to the symmetry term
of the effective interaction {\it below saturation density}
\cite{cdl98}, thus providing a good opportunity to distinguish between
different isospin EOS.

In equilibrium statistical multifragmentation
calculations \cite{smm95} sensitive observables 
appear to be the yield ratios of light isobars (e.g. $t/^3He$)
and the formation of more stable primary intermediate mass 
fragments \cite{bot}. Therefore
in the following, from "ab initio" dynamical simulations
of central collisions, we will discuss the effects of 
the structure of the instabilities 
on fragment observables (mass/charge-yields, isospin content and
multiplicity distributions) and on the reaction mechanism, i.e. the
time evolution of the isospin dynamics. 
Complementary 
information will be obtained from the comparison of bulk and
neck fragmentation events in semi-central and semi-peripheral collisions, 
where the latter represent a large
part of the reaction cross section \cite{neck}. 

\section{Stochastic Transport Simulations}

A new code for the solution of microscopic transport equations
of Boltzmann-Nordheim-Vlasov ($BNV$) type
\cite{ber88,gre87,bon94,bon91,gua96}
has been written where asymmetry effects are
suitably accounted for \cite{fab98,flows} and where fluctuations 
are included dynamically \cite{flu98}. A density dependent symmetry term is
used  consistently in the construction of the initial
ground state and the evolution of the reaction, isospin effects on 
nucleon cross sections and
Pauli blocking are consistently evaluated.
The transport equations are solved following a test particle evolution on 
a lattice \cite{gua96,gre98}.
A parametrization of free NN cross sections is used with
energy and angular dependence.

It is important to include fluctuations when discussing dynamical
fragment formation.
Indeed, the evolution under the influence of 
fluctuations is described by a transport equation with a 
stochastic fluctuating term, the so-called 
Boltz\-mann-Lan\-ge\-vin equation
\cite{Sakir,jorgen,BOB}. 
In this paper we will follow two methods to 
include fluctuations, the Stochastic Mean Field method\cite{flu98}
and the corresponding simplified approach used in 
\cite{noise,box94,alfio}.
The Stochastic Mean Field method is based on a fully self-consistent
treatment of the fluctuations during the time evolution.
On the path towards local thermal equilibrium, 
the system is characterized by a mean trajectory $\bar f$ 
and a variance $\sigma^2 = <(f-\bar f)^2>$ 
which in each phase space cell obeys the equation of motion
\begin{equation}
\frac{d}{dt}\sigma^2 = -\frac{2}{\tau(t)}\sigma^2 + 2~D(t).
\label{sigm}
\end{equation}
with $2~D(t)$ the correlation function of the fluctuating term
 and $\tau(t) = 1 / (w^++w^-)$, 
where $w^+$ and $w^-$ are the transition probabilities 
into and out of the phase space cell. 
The equilibrium statistical 
value $\sigma_0^2 = f_0(t_{eq})(1-f_0(t_{eq}))$ 
suggests an ansatz for the correlation 
function of the fluctuation term of the form
\begin{equation}
2D(t) = (1-\bar f) w^+ + \bar f w^-
\label{corr}
\end{equation}
i.e. the magnitude of the  fluctuations is given by the 
total number of collisions  
(fluctuation-dissipation theorem) \cite{jorgen}.  
Then we obtain for the time evolution of the quantity
$\Delta = \sigma^2 - \bar f (1- \bar f)$
\begin{equation}
\frac{d}{dt}\Delta = -\frac{2}{\tau(t)}\Delta .
\label{diff}
\end{equation}
$\Delta = 0$ is a solution of eq.(\ref{diff}).
Therefore, if the variance is initially locally set equal to its 
statistical value, it  
will always be given by the local statistical variance 
as  $\sigma^2 =  \bar f (1- \bar f)$.
With a projection of this relation on coordinate space we 
obtain  local density
fluctuations which are implemented with a Monte-Carlo method
at each time step. In this way we also have a branching of 
trajectories.
The procedure is valid if we assume a local statistical 
equilibrium, appropriate for the problems discussed here,
i.e. fragment production in the expansion/separation phase.

The other approach, computationally much easier \cite{noise,box94,alfio}, 
is based on the introduction of density fluctuations by a random sampling
of the phase space. The amplitude of the noise is
gauged to reproduce the dynamics of the most 
unstable modes \cite{box94}. 
For each system we have checked the equivalence of the two methods
in the description of the collision dynamics
in the complete evolution  from
fast particle emissions to the fragment production.
The analysis of the results, presented in the next Section, is
based on events collected with both methods. In the 
implementations isospin effects in the fluctuations have been 
consistently accounted for. 

\section{Dynamical Simulations} 

In this section
we want to show the possibility of extracting quantitative information
on the symmetry term of the EOS directly
from fragmentation reactions using stable isotopes with different 
charge asymmetries. The first exclusive data for such
reactions are now starting to become available 
\cite{xu00,vercr,ts01,yencr,schcr,chimera,desouza}. 
We study neutron-rich and neutron-poor $Sn + Sn$ collisions.
In spite of the relatively low asymmetries tested
with stable isotopes we will see an interesting
and promising dependence on the stiffness of the symmetry term.

The dynamics of a fragmentation reaction is extremely rich.
The time evolution can be roughly divided into the following
phases: pre-equilibrium particle emission, compression-expansion stage,
multifragmentation ending in the freeze-out configuration, and 
statistical decay of the  primary fragments. 
All the steps are seen to be very isospin dependent
and therefore it is essential to perform consistent "ab initio"
simulations of the whole process in order to extract 
information on the symmetry term of the EOS. 
Since in the reaction dynamics with intermediate energy
beams the asymmetric nuclear matter probes compressed as
well as dilute phases the final output will be determined by the
complete density dependence of symmetry term. The aim of this paper 
is to show which fragmentation observables
are particularly sensitive to the symmetry
term of the EOS. We will see that the physical interpretation of the results 
can be made quite transparent in spite of the complexity of the numerical 
simulations. 
The new features of the liquid-gas
phase transition in asymmetric nuclear systems discussed in the
previous sections were based
on thermodynamical considerations \cite{mue95,bao197,bar01},
linear response approaches \cite{cdl98} or dynamical simulations in
a box \cite{bar98} (i.e. without finite size and Coulomb effects). 
It is important to check these predictions in
realistic simulations of fragmentation
reactions for systems with different charge asymmetries.

In order to simplify the analysis of the most sensitive observables
to isospin effects we have chosen a Skyrme force with
the same {\it soft} EOS for symmetric NM ($K=201MeV$)
and with three different choices for the density dependence of the symmetry
term, i.e. the asy-soft, stiff,  
and superstiff interactions discussed in sect. 2, and 
qualitatively seen in fig. 2. 
In this way we force the symmetric part of the EOS to be exactly 
the same in order to disentangle dynamical symmetry term effects. 
As we see in fig. 2  the potential symmetry term 
for the various  interactions shows quite different behaviours 
in the region
around normal density and at very low densities. 
While around $\rho_0$ the density dependence becomes steeper
when we go from asy-soft to asy-superstiff (this suggests the names)
at subsaturation densities, where we enter
the spinodal zone and the fragment formation initiates, 
they manifest an opposite trend.  
We will show that the reaction mechanism is sensitive to these
different behaviours, even though with stable 
nuclei we are limited in the
possible asymmetries to be explored (and moreover we will 
certainly not reach high compression
regions). 

\subsection {Reaction Mechanisms}

We have studied  collisions of the systems 
$^{124}Sn+^{124}Sn$ and $^{112}Sn+^{112}Sn$ at $50AMeV$, 
where new data
have just appeared or are under analysis at MSU 
\cite{xu00,vercr,ts01}.
We investigate  semi-central ($b=2fm$) 
and semi-peripheral ($b=6fm$) reactions for each of the three 
isospin EOS. The analyses are based on  around 500 events
in each case. In order to get a qualitative impression of 
the reaction mechanism the time evolution of the
density (projected on the reaction plane) is shown 
in figs. 5 and 6
for impact parameters $b=2,4,$ and $6 fm$ 
 for one typical event each (neutron rich case,
asy-stiff EOS). 

For a semi-central collision, $b=2fm$, the reaction mechanism 
corresponds to bulk fragmentation. We can identify three
main stages that are characterized by 
specific features of the isospin
dynamics since the system explores different density
regions. After a first compression phase 
(until about 40-50 $fm/c$) a fast expansion phase follows
(until 110-120 $fm/c$). Then during the fragmentation stage 
the system will break up and the fragment
formation process takes place up to the {\it freeze-out} time
(around 260-280 $fm/c$). At this time excited primary fragments have been
formed which are far enough apart to have
a negligible mutual nuclear interaction (see fig.4). 
This freeze-out time is relatively well defined in the simulations 
as the time when
the average number of produced fragments stabilizes.
The physical conditions of density and
temperature at the beginning of the fragmentation  stage 
correspond to an unstable
nuclear matter phase. The volume instabilities have time to develop and
we expect to see a spinodal decomposition  
with the formation of a liquid and a gas phase.

For $b=6fm$, (semi-peripheral collision, fig. 6) we observe quite a different
behaviour.  Now in the overlap region 
a neck structure is developing. During the 
interaction time (from about 80 to 120 $fm/c$) it heats and expands 
but remains  in contact  with the denser and colder regions 
of the projectile-like 
(PLF) and/or target-like (TLF) fragments. 
Now the surface instabilities of a
cylindrically shaped neck region and the
fast leading  motion of the PLF and TLF will play the
important role. At the freeze-out time with the neck rupture 
at about 140 $fm/c$ 
intermediate mass fragments (IMF) are produced
in the mid-rapidity zone. One can have a large variety
of event structures, typical of a dynamical instability: in some events 
fragments are formed very early or, in others, they can remain for a 
longer time attached to the leading PLF or TLF fragments 
\cite{neck,tok96,dur98,indra}.
Moreover, the projectile- and target-like primary fragments can
be quite deformed at the rupture time and may follow a fast-fission 
path of purely dynamical nature \cite{col00}. For a given 
semi-peripheral impact parameter we expect to see
a very wide velocity distribution in the fragment sources \cite{neck}.

An intermediate behaviour between bulk and neck fragmentation, 
is observed for $b=4fm$ (fig.6).
The freeze-out time is decreasing with impact parameter. 
This gradual transition suggests that it is quite
inapproppriate to discuss all events in terms of a
unique fragmentation mechanism and even harder to try to
assign a fixed size or shape to a multifragmenting source,
even passing from $b=2 fm$ to $b= 6 fm$. Moreover the prompt
nucleon emission should be always accounted for, at each step
of the reaction dynamics.

\begin{figure}[htb]
\begin{minipage}[t]{67mm}
\epsfysize=11.cm
\vspace{5cm}
\centerline{\epsfbox{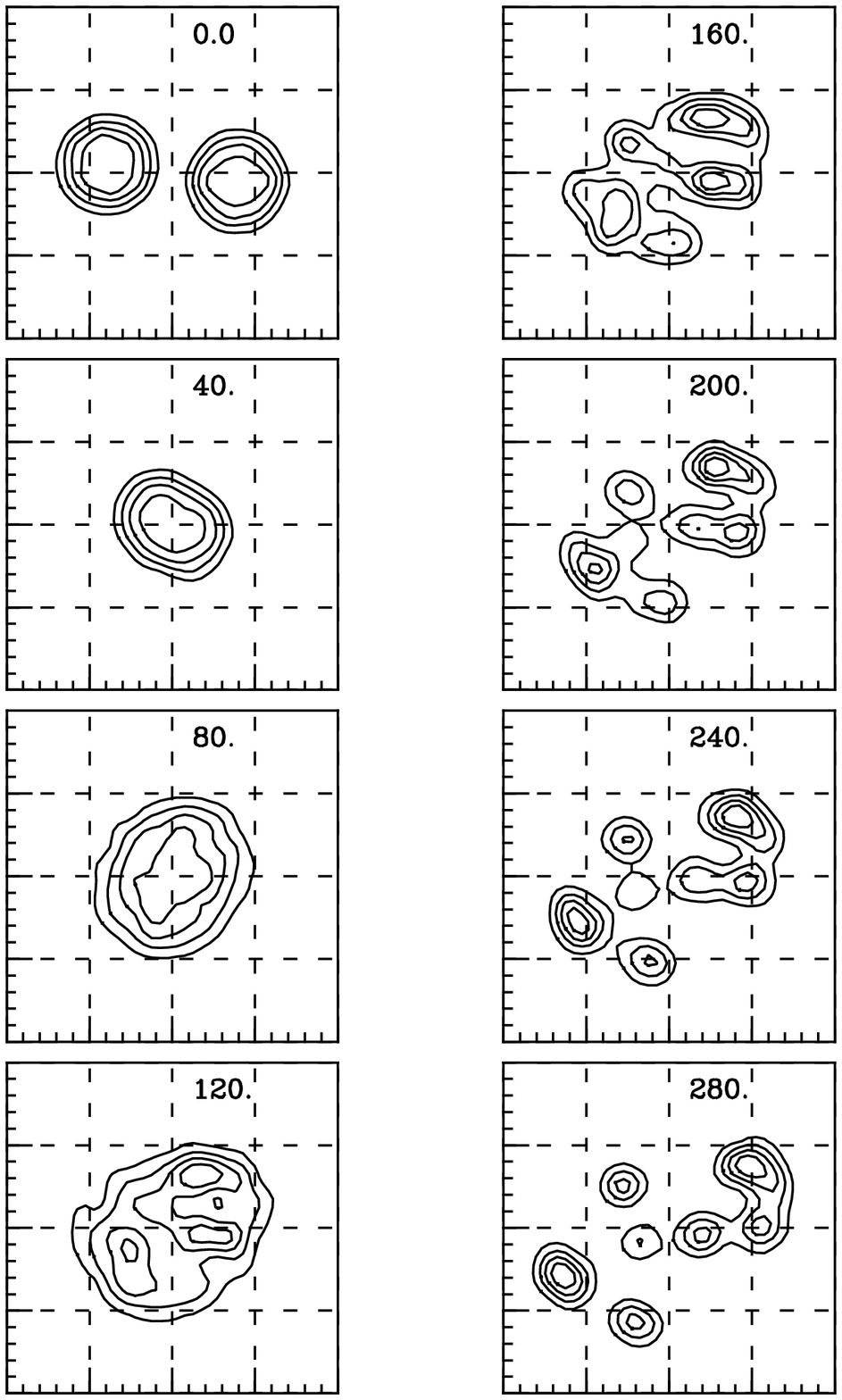}}
\caption{\it $^{124}Sn+^{124}Sn$ collision at $50AMeV$: time evolution of the
nucleon density projected on the reaction plane.
Semi-central $b=2fm$ collision: approaching, compression, separation, and
fragmentation phases. The times in $fm/c$ are written in each panel.
The iso-density lines are plotted every $0.02fm^{-3}$
starting from $0.02fm^{-3}$.}
\end{minipage}
\hspace{\fill}
\begin{minipage}[t]{67mm}
\epsfysize=11.cm
\vspace{5cm}
\centerline{\epsfbox{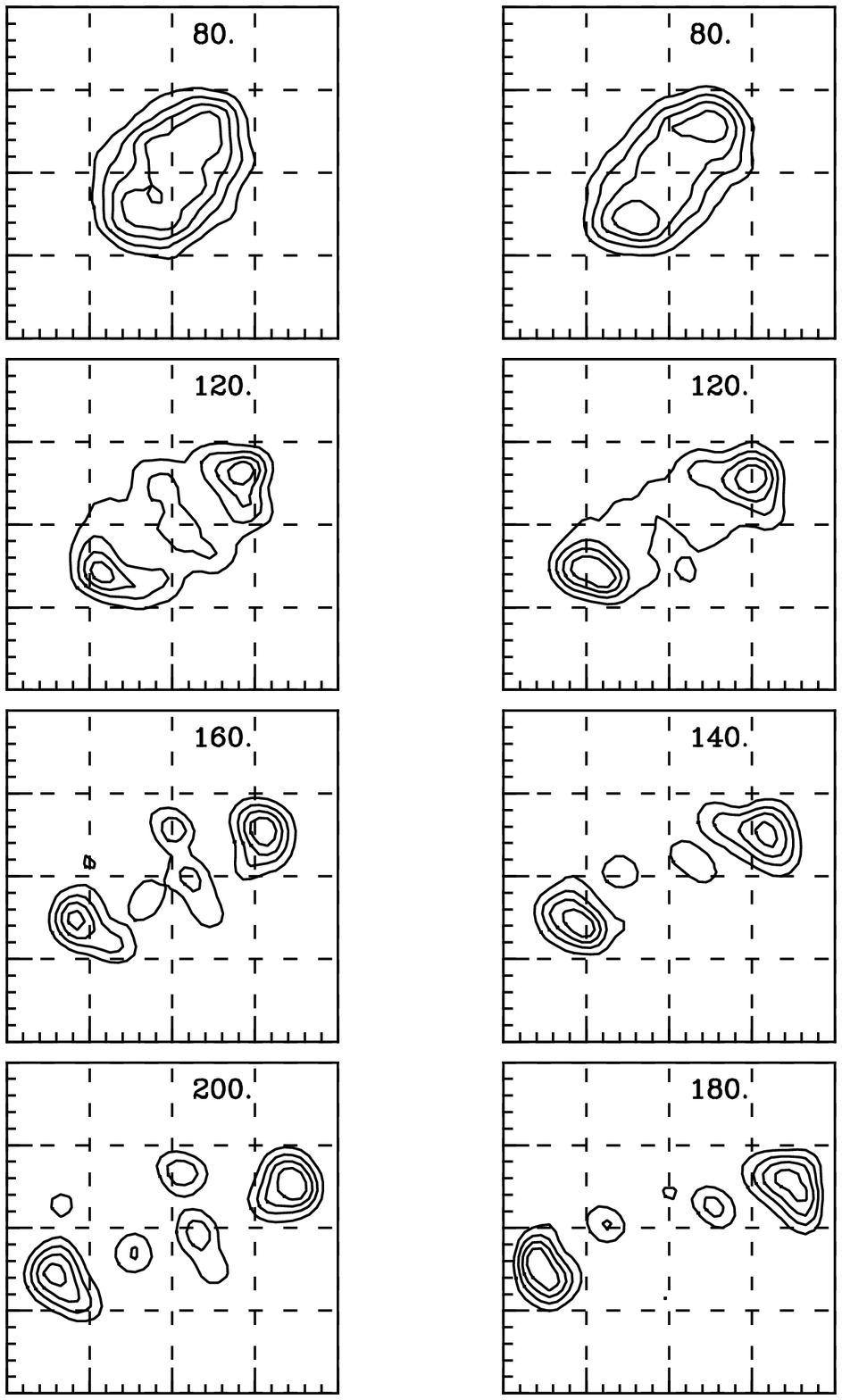}}
\caption{\it $^{124}Sn+^{124}Sn$ collision at $50AMeV$: time evolution of the
nucleon density projected on the reaction plane, like in previous figure.
First column: $b=4fm$ , second column: $b=6fm$,
separation phase up to the freeze-out.}
\end{minipage}
\end{figure}

A different aspect of the reaction mechanisms can be also seen 
from fig.7
where we show the parallel velocity distribution (in the $CM$ frame)
of the produced fragments ($Z \geq 3$) for semi-central (fig.7a),
and semi-peripheral (fig.7b) collisions (neutron-rich case, asy-stiff EOS).
In the semi-peripheral reactions (panel b) one can clearly distinguish 
the contributions of the spectators and the neck-IMF's. The slope 
of the spectator contributions is due to displaying the velocity 
rather than the momentum. A
remnant of the spectator contribution is also seen in the 
semi-central events for $b = 2 fm$.   
Of particular interest is the larger velocity spread of the
IMF's in the semi-peripheral relative to the semi-central 
events, which points to a different production mechanism. In general,
the investigation of velocity distributions and correlations should
yield further insight into the mechanism of the reactions, but
we defer this to later work. 

\begin{figure}[htp]
\epsfysize=6.cm
\vspace{5cm}
\centerline{\epsfbox{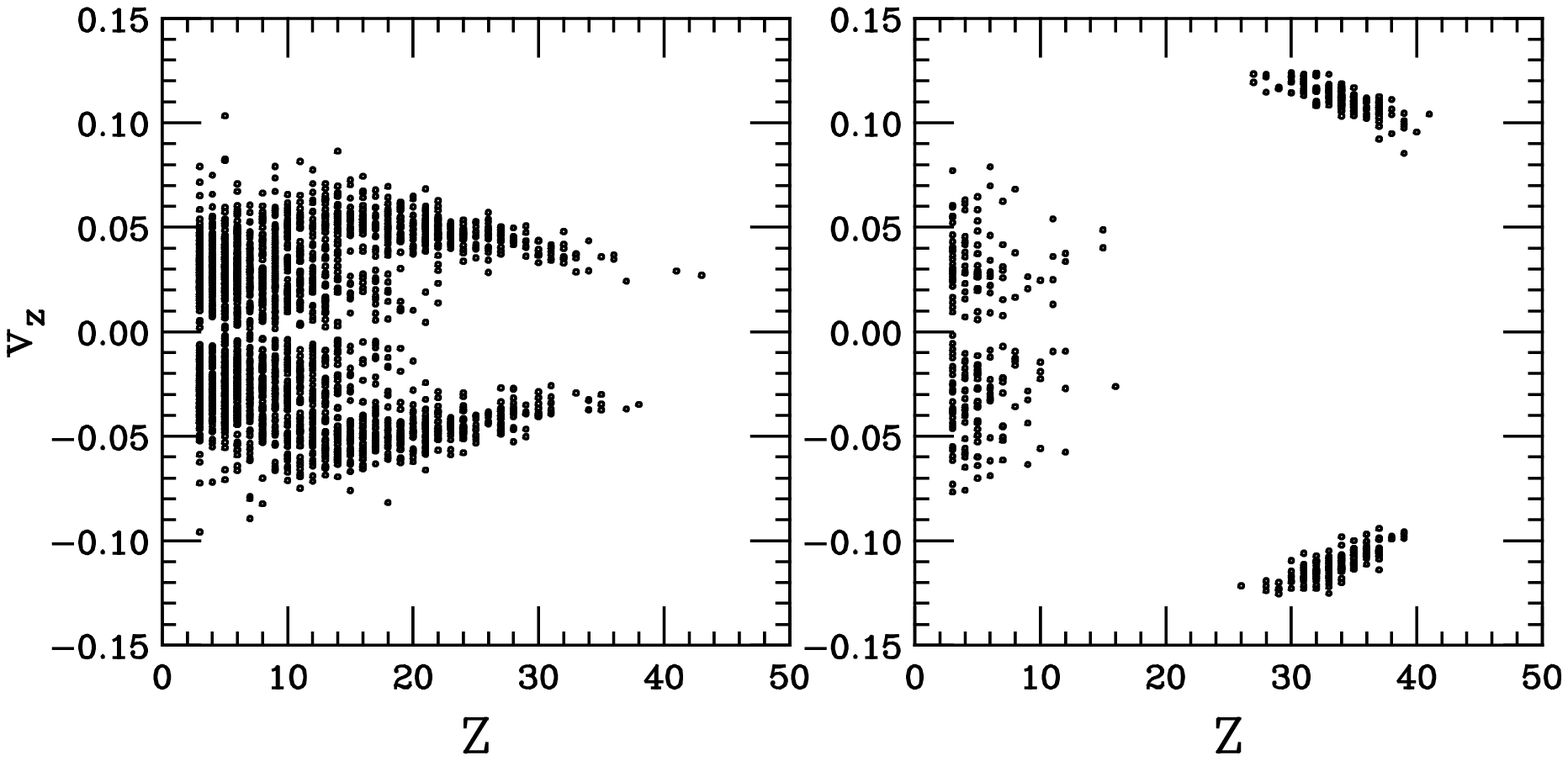}}
\caption
{\it $^{124}Sn+^{124}Sn$ collision at $50AMeV$: parallel velocity
distribution of all fragments ($Z \geq 3$) at freeze-out time
for an asy-stiff EOS:
a) semi-central events ($b=2fm$); b) Semi-peripheral events ($b=6fm$).
Asystiff EOS.
}
\end{figure}

Guided by the density contour plots we have performed in the following a
separate analysis of some quantities in a central region having a 
linear dimension of $20fm$ centered in the $CM$ of the system. 
 Since this corresponds to the active
volume in which fragmentation takes place (see figs. 5,6) we obtain
in this way a more
detailed picture of this process.

We organize the presentation of the results of the simulations 
in the following way. We first analyse the case of the $n$-rich
reaction $^{124}Sn+^{124}Sn$. We study for the asy-stiff 
interaction the reaction mechanism for semi-central and semi-peripheral 
collisions. Then we discuss the effects of different assumptions on the
symmetry term. After that we discuss the differences for the $n$-poor case 
$^{112}Sn+^{112}Sn$. 

\subsection{Results of $^{124}Sn+^{124}Sn$, $n$-rich case}

The results obtained with the
{\it asy-stiff} symmetry term are shown in fig.8 (semi-central, $b=2fm$)
and fig.9 (semi-peripheral, $b=6fm$). For each reaction the results 
are presented here and in the following cases in the following way:

Left column, time evolution of: (a) {\it Mass} in the
liquid phase, $Z\geq3$ (solid line and dots) and the gas phase 
(solid line and squares); 
(b) {\it Asymmetry} $I=(N-Z)/(N+Z)$ in
the gas "central" (solid line and squares), gas total (dashed and squares), 
liquid "central" (solid and dots) phase and for IMF ($3<Z<15$, stars). 
The horizontal line shows the initial
asymmetry; 
(c) {\it Mean Fragment Multiplicity} $Z\geq3$.
The saturation of this curve defines the freeze-out time and 
configuration.
Right column, distributions of the "primary" fragments in the
{\it freeze-out configuration}: (d) {\it Charge distribution}, 
 (e) {\it Asymmetry distribution} as a function of the 
fragment charge and
(f) {\it Fragment multiplicity distribution} (normalized to $1$).

\begin{figure}[htb]
\begin{minipage}[t]{67mm}
\epsfysize=9.75cm
\vspace{5cm}
\centerline{\epsfbox{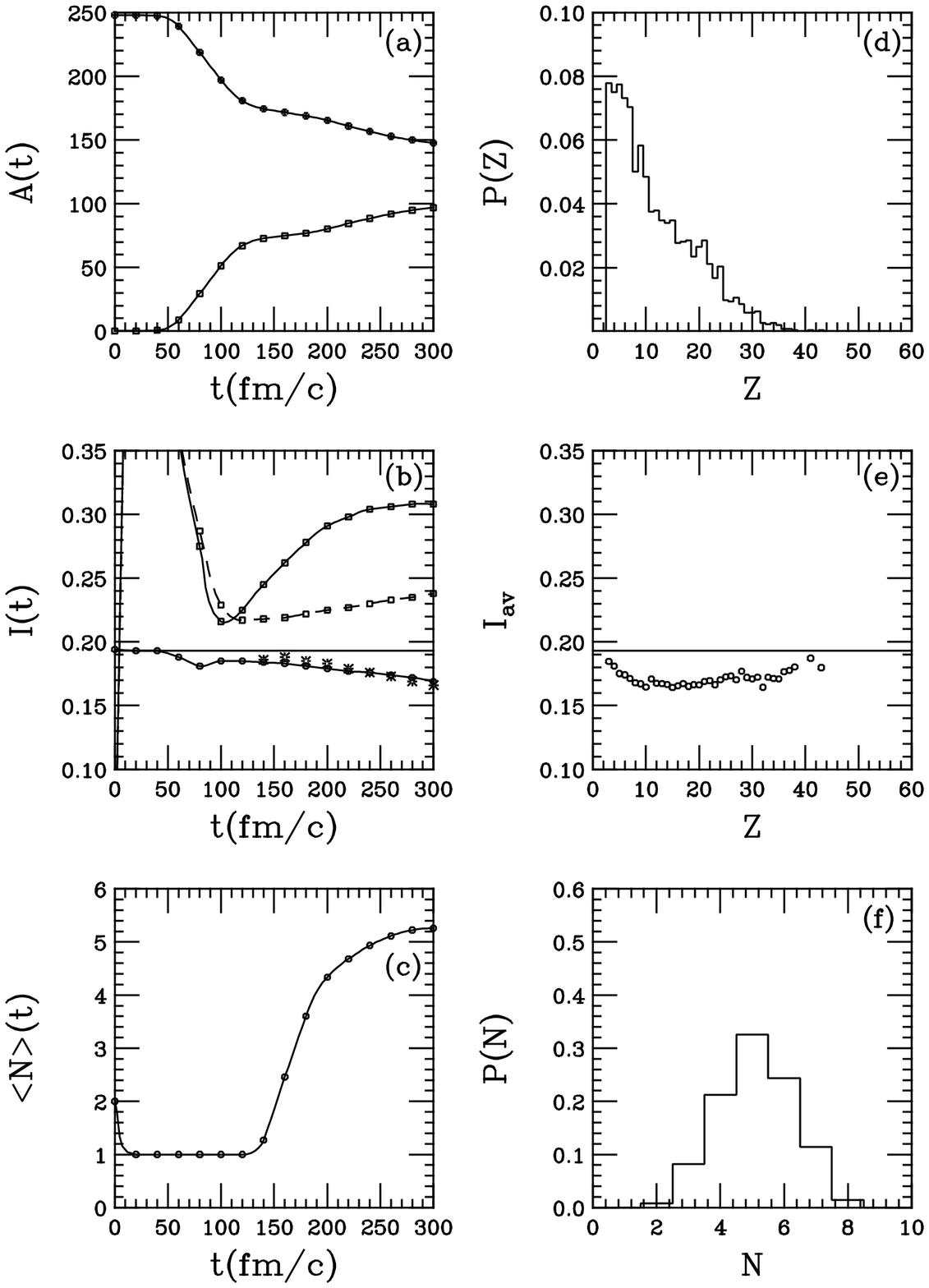}}
\caption{\it $^{124}Sn+^{124}Sn$ $b=2fm$ collision: time
evolution (left) and freeze-out distributions (right), 
see text (ASY-STIFF EOS).}
\end{minipage}
\hspace{\fill}
\begin{minipage}[t]{67mm}
\epsfysize=9.75cm
\vspace{5cm}
\centerline{\epsfbox{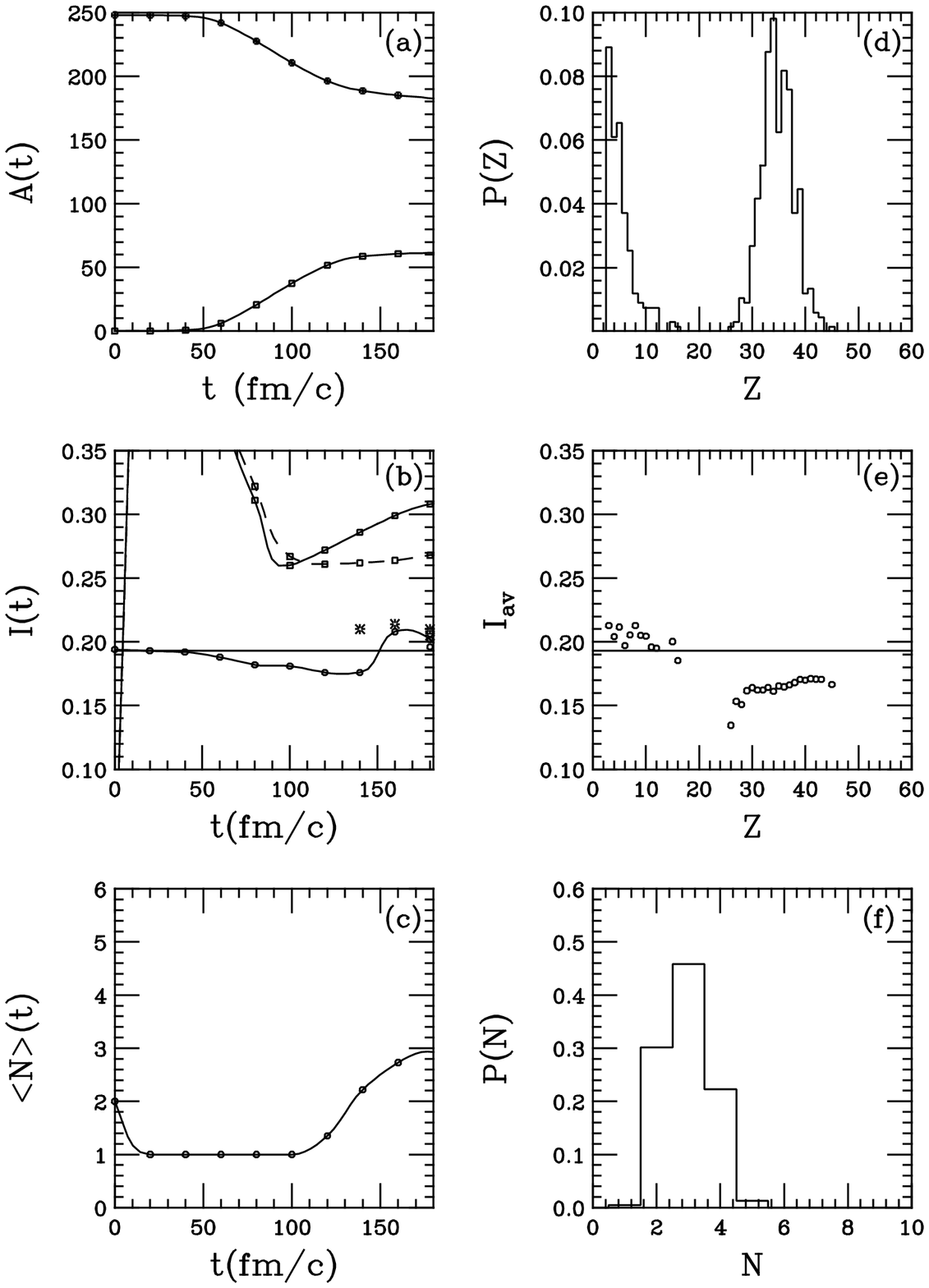}}
\caption{\it $^{124}Sn+^{124}Sn$ $b=6fm$ collision: time
evolution (left) and freeze-out distributions (right), 
see text. ASY-STIFF EOS.}
\end{minipage}
\end{figure}

From these figures we can obtain a clear picture of the reaction 
mechanism. E.g. looking at fig. 8a we see the approach and 
compression phase up to about 50 $fm/c$ in which almost all mass 
is still in the 
liquid, except a few pre-equilibrium particles. Between 50 and 
120 $fm/c$ we have the expansion phase in which many particles are 
emitted into the gas, but there is still one fragment (fig. 8c).
In the fragmentation phase between 120 and 280 $fm/c$ the number 
of fragments rises rapidly (fig. 8c), but fewer particles are emitted into 
the gas. The freeze-out time is characterized by the saturation of
the number of fragments in fig. 8c. The same reaction phases are seen in 
the semi-peripheral collision (figs. 9a,c), however, somewhat less destinct.
The final charge distribution of semi-central and semi-peripheral collisions 
(figs. 8d and 9d) are
distinctly different: in more central collisions it is rapidly decreasing, 
while in more peripheral collisions it shows a second maximum at the 
average charge of the spectators. Here the light fragments below
charge 10 are the IMF produced in the neck. 
The average number of produced 
fragments is higher in the central case (figs. 8c and 9c at freeze-out 
times) and correspondingly  the multiplicity of fragments is 
higher (figs. 8f and 9f). For the study of 
isospin dynamics the
panels $b$ and $e$ are the most interesting, which give the evolution 
and final distribution of the isospin content of the various 
components in the reaction. They will be discussed below for the different
reaction types.

\subsubsection {Semi-central collisions: bulk fragmentation.}
 
We now discuss the isospin dynamics in the semi-central collisions,
i.e. the behaviour given in figs. 8b,e. As seen the  
pre-equilibrium particles emitted during 
the compression phase are predominantly neutrons, as expected for
the neutron-rich system. The particles emitted into the gas 
during the expansion phase consist equally of protons and neutron, 
thus lowering the asymmetry of the gas.
The liquid phase becomes more symmetric during the compression
due to neutron emission and then remains essentially constant 
during the expansion.
At the beginning of the fragmentation  phase, when density 
inhomogenities start
to develop, an {\it isospin burst} of gas phase
in the "central region" is observed (fig.8b) . At the same time
the central liquid phase is becoming more and more
symmetric. This behaviour is consistent with the dynamical spinodal 
mechanism in dilute asymmetric nuclear matter leading to isospin 
distillation.

In addition we observe very interesting features in 
the isospin content of the
primary fragments in fig. 8e. We clearly see the 
isospin distillation
effect, i.e. all IMF's are formed more symmetrically 
than the initial asymmetry.
Moreover in fig.8e we can distinguish two opposite
trends  for fragments with charge above and below roughly $Z=15$.
For the heavier fragments   
the average asymmetry increases with the charge. This is expected 
since for heavier nuclei a larger Coulomb effect must be 
compensated . However, the asymmetry rises again towards 
lighter fragments contrary to the trend for stable isotopes.
We interpret this behaviour as 
a result of the different density regions in which 
fragments form and grow. 
The trend 
can be understood in a dynamical scheme following
the transport effects of the symmetry term.
This so-called {\it proton migration} effect, that yields more
symmetric larger fragments, is seen more clearly
in the more peripheral collisions, and will be discussed there. 

However, as shown also in the 
density contour plot (fig.5) not all
fragments form simultaneously, that makes the interpretation of the results
not trivial. 

\subsubsection {Semi-peripheral collisions: neck fragmentation.}

For semi-peripheral collisions at $b=6fm$ the isospin content 
appears with some new distinctive features as seen in fig. 9e. 
The IMF fragments
formed in the neck region are much more asymmetric (more neutron rich)
than the corresponding
fragments produced in semi-central collisions (cf. fig. 8e). 
The heavy PLF and TLF have a
definite lower asymmetry than the IMF's. 
We believe this to be due to  
a different nature of the fragmentation mechanism in the neck
region, indicating a transition from volume to shape instabilities with
a different isospin dynamics.
To interpret this  we have to keep
in mind the following features of "neck-fragmentation":
(i) The clusters are formed in a nuclear matter not very dilute
relative to saturation density. We expect 
some density fluctuations but the fragmentation is mainly due to
the shape dynamics \cite{neck}.
(ii) The neck region is always in contact with a "high" density
phase (the spectators) during the fragment formation.
(iii) Due to a sponateous symmetry breaking of the neck instability  
in some events the fragments are formed closer to
one of the  spectators with increased interaction betweeen the
two. Thus there will be a  smooth transition to PLF/TLF 
fast-fission type of events \cite{col00}.

Due to the increase of the symmetry term  just below normal density
it will be energetically more favorable to {\it migrate} protons from
the neck region to the more dense spectators, leaving the nuclear matter
in the neck more neutron rich at the time of breaking.
In a sense the isospin dynamics is ruled by the same energetic argument
as in the case of isospin distillation in central collisions. 
The main difference to the case of bulk fragmentation is now, that 
there is a spatial separation of the neutron-rich neck and the spectators,
which initiates a flow of protons resp. neutrons in opposite directions.
It thus involves an isovector mode of the mean field.
Therefore we propose to call this phenomenon {\it isospin migration} in 
contradistinction to the distillation phenomenon, which is a 
spontaneous separation of the two phases. 

Another 
difference is that we are testing the symmetry energy in different
regions of nucleon density. In central collisions  fragments are produced
with the chemical instability mechanism in a very dilute asymmetric nuclear 
matter. The neutron distillation is therefore due to the symmetry energy
increase at very low densities between $0.03~and~0.10fm^{-3}$ 
in our simulations. The
neck matter, on the other hand, ruptures starting from 
densities just below the saturation
values in contact with a normal density region given by the spectators.
The symmetry energy range of interest here is now still in the
subnuclear range but closer to $\rho_0$, between $0.08~and~0.16fm^{-3}$
in our simulations.
This interpretation of the difference in the average isospin content 
of the fragments
produced in semi-central and semi-peripheral collisions has a promising
aspect: The sensitivity to the density dependence of the
symmetry term in different regions of the subnuclear density range
would allow the possibility of distinguishing between different
effective interactions. This effect will be clearly seen below in 
our simulations.

The rise of the asymmetry of the IMF in semi-central collisions 
(fig. 8e) can also be interpreted as an isospin migration effect.
In bulk fragmentation the smaller IMF's are also formed in contact 
with bigger fragments in the late stages of the process. Thus 
their isospin asymmetry increases by the same mechanism 
as that of neck fragments in peripheral colliions. 

\subsubsection {Symmetry term effects}

In this section we investigate the effects of different assumptions 
on the density dependence of the symmetry energy.
In figs. 10,11 and 12,13 we report the results, 
obtained respectively with the {\it asy-soft}
and the {\it asy-superstiff} symmetry term in the same format as was 
done for the {\it asy-stiff} symmetry term in figs. 8,9. The results 
for semi-central collision events, $b=2fm$, are thus in the figs. 8,10, and 12,
while those for semi-peripheral collisions, $b=6fm$, are shown 
in figs. 9,11, and 13.

As already observed the reaction dynamics is very
rich. Thus also the fast particle emission in the 
expansion phase is affected by the symmetry term, in particular
in the $N/Z$ composition. As seen before a large number of
nucleons are emitted
into the gas in the time interval between 50 and 120 $fm/c$,
i.e. in the expansion phase (see fig.5). This means that the
symmetry part of the mean field at subnuclear densities will
have an important dynamical effect on the emission. In the asy-soft
case below $\rho_0$ neutrons are less bound than in the asy-superstiff
case,
and oppositely for the protons: we then expect a more neutron-rich
prompt particle emission with the asy-soft symmetry term and 
a more symmetric initial dilute matter which will undergo fragment 
production. This is exactly what we see from the figs. 10b (asy-soft)
 and fig.12b (asy-superstiff) for central collisions. As expected the
results of fig.8b (asy-stiff) are somewhat in between. It is 
interesting 
to remark that the same effect is present also for semi-peripheral
collisions (compare figs. 11b and 13b), although more weakly.

\begin{figure}[htp]
\begin{minipage}[t]{67mm}
\epsfysize=9.75cm
\vspace{6cm}
\centerline{\epsfbox{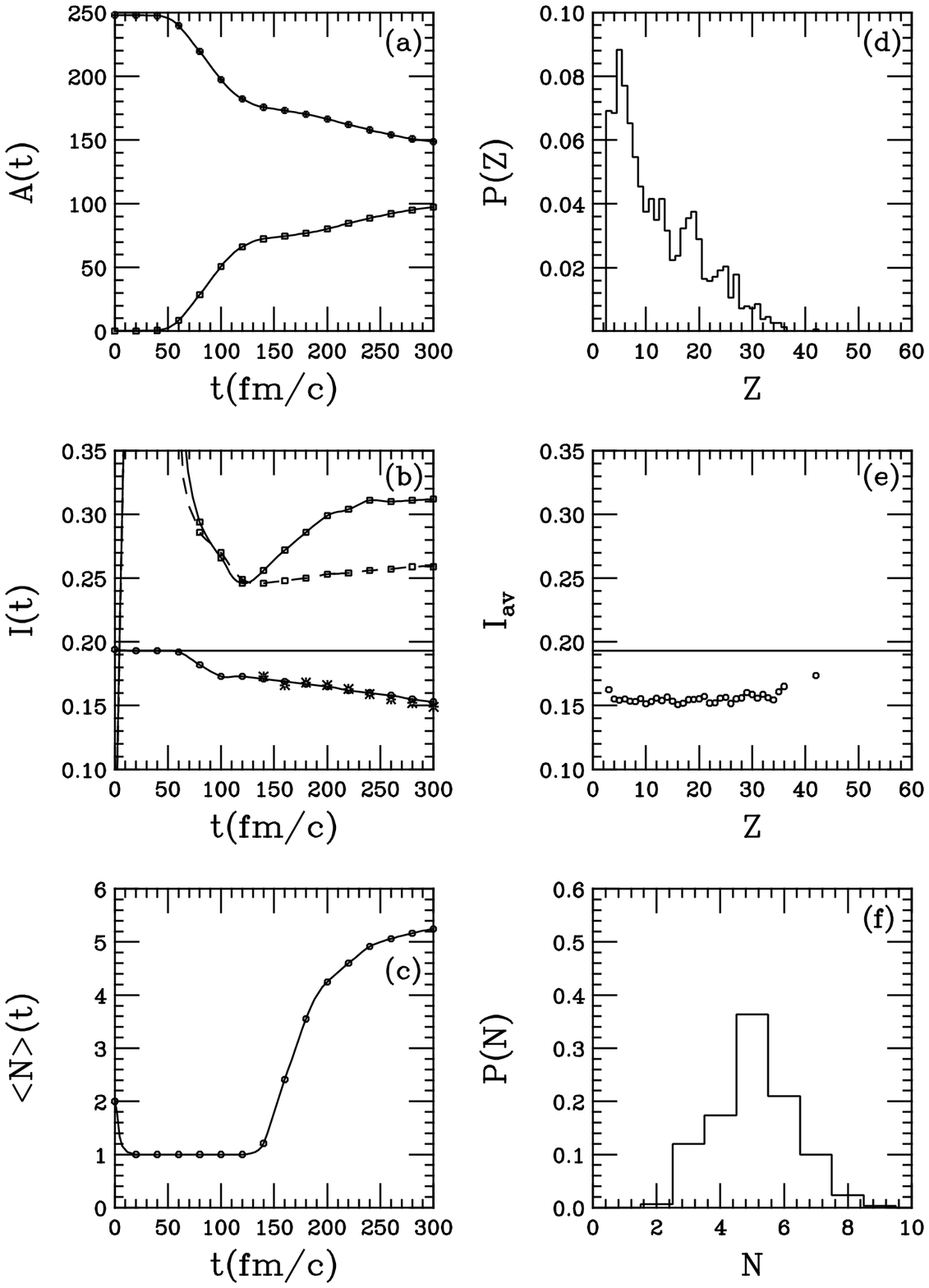}}
\caption{\it $^{124}Sn+^{124}Sn$ $b=2fm$ collision: time
evolution (left) and freeze-out properties (right).
See text. ASY-SOFT EOS}
\end{minipage}
\hspace{\fill}
\begin{minipage}[t]{67mm}
\epsfysize=9.75cm
\vspace{6cm}
\centerline{\epsfbox{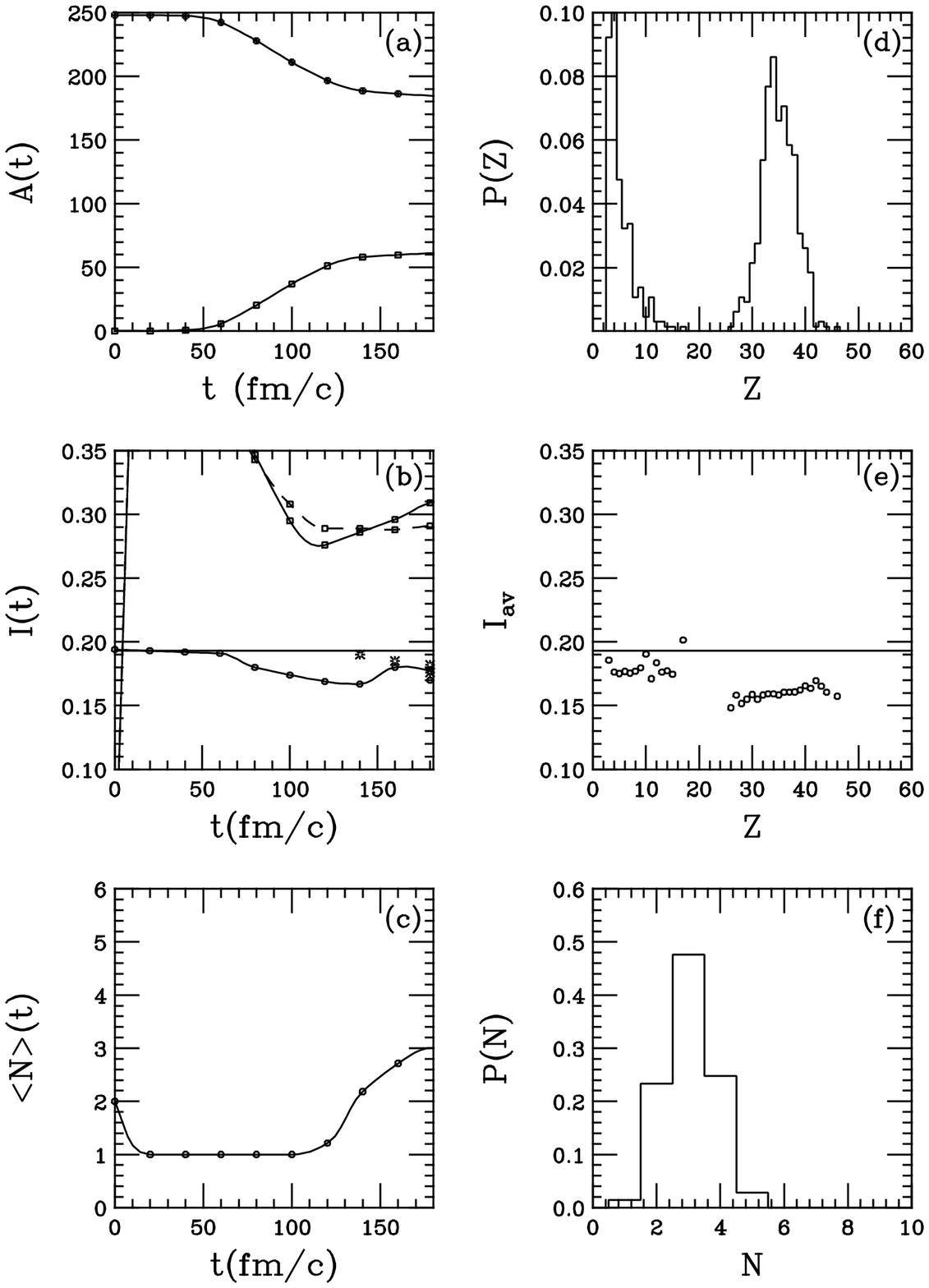}}
\caption{\it $^{124}Sn+^{124}Sn$ $b=6fm$ collision: time
evolution (left) and freeze-out properties (right).
See text. ASY-SOFT EOS.}
\end{minipage}
\end{figure}

We will now focus our discussion on the isospin content
of the primary intermediate mass fragments ($3 \leq Z \leq 20$).
From a comparison of figs. 8e,10e and 12e for semi-central events
and of figs. 9e,11e and 13e for semi-peripheral events we clearly see
that the asysoft choice is the most effective for the neutron
distillation effect (the most symmetric IMF's) in semi-central collisions
while the asy-superstiff choice is the most effective in forming
neutron-rich IMF's in the neck region for semi-peripheral collisions.
Following the above discussion we easily understand these two
trends considering the differences in the symmetry term
at subnuclear densities (see figs. 1 and 2 of sect.2).
The asy-soft parametrization gives a larger symmetry energy 
and a rather flat behaviour below normal density.
Thus it is quite efficient during the chemical instability growth
but not so for the proton migration from neck
to spectators. Exactly the opposite trend is seen for the asy-superstiff
parametrization with roughly a quadratic density dependence 
and thus a weak increase
at low densities and a steep slope around $\rho_0$.

\begin{figure}
\begin{minipage}[t]{67mm}
\epsfysize=9.75cm
\vspace{6cm}
\centerline{\epsfbox{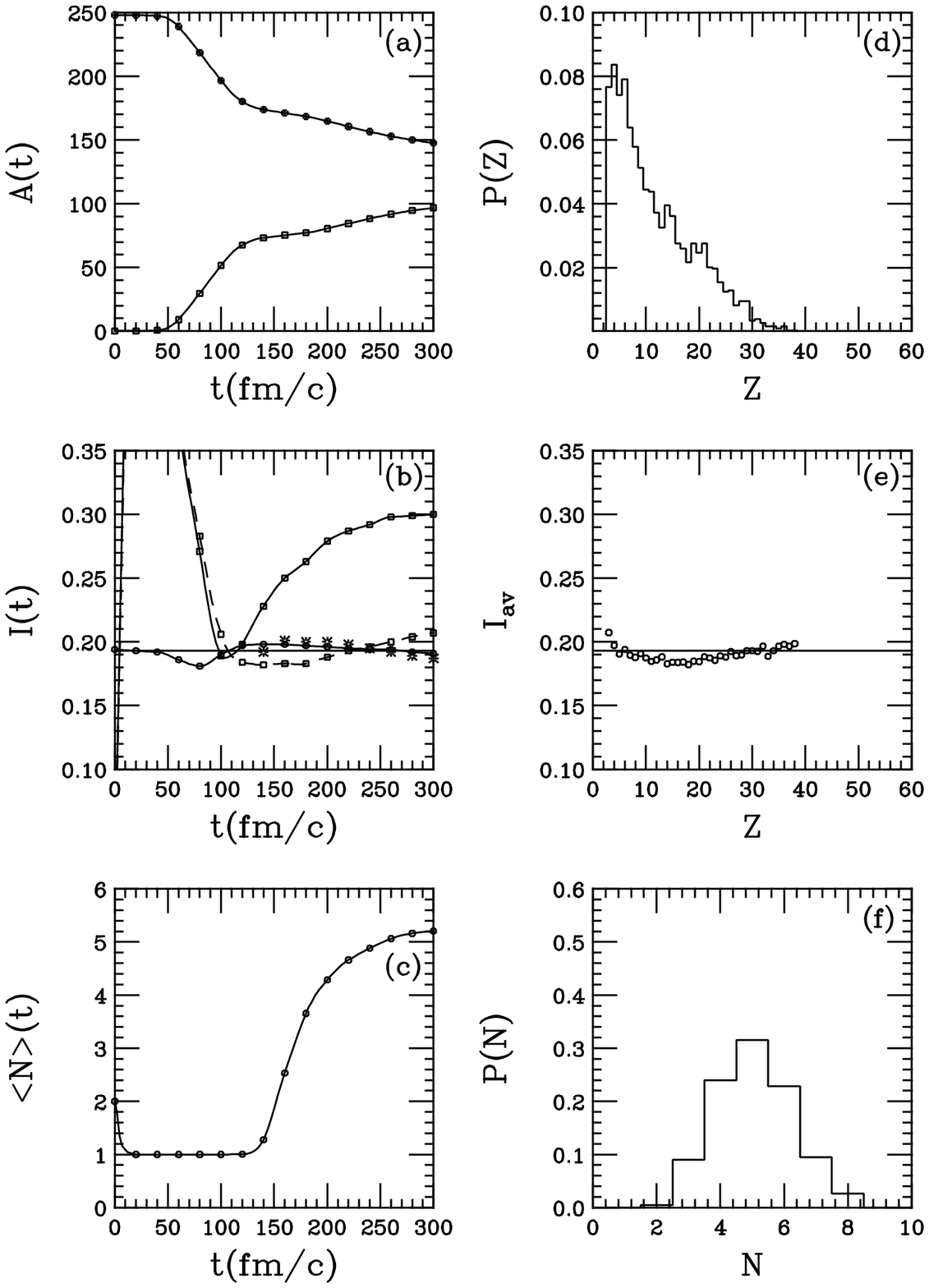}}
\caption{\it $^{124}Sn+^{124}Sn$ $b=2fm$ collision: time
evolution (left) and freeze-out properties (right).
See text. ASY-SUPERSTIFF EOS}
\end{minipage}
\hspace{\fill}
\begin{minipage}[t]{67mm}
\epsfysize=9.75cm
\vspace{6cm}
\centerline{\epsfbox{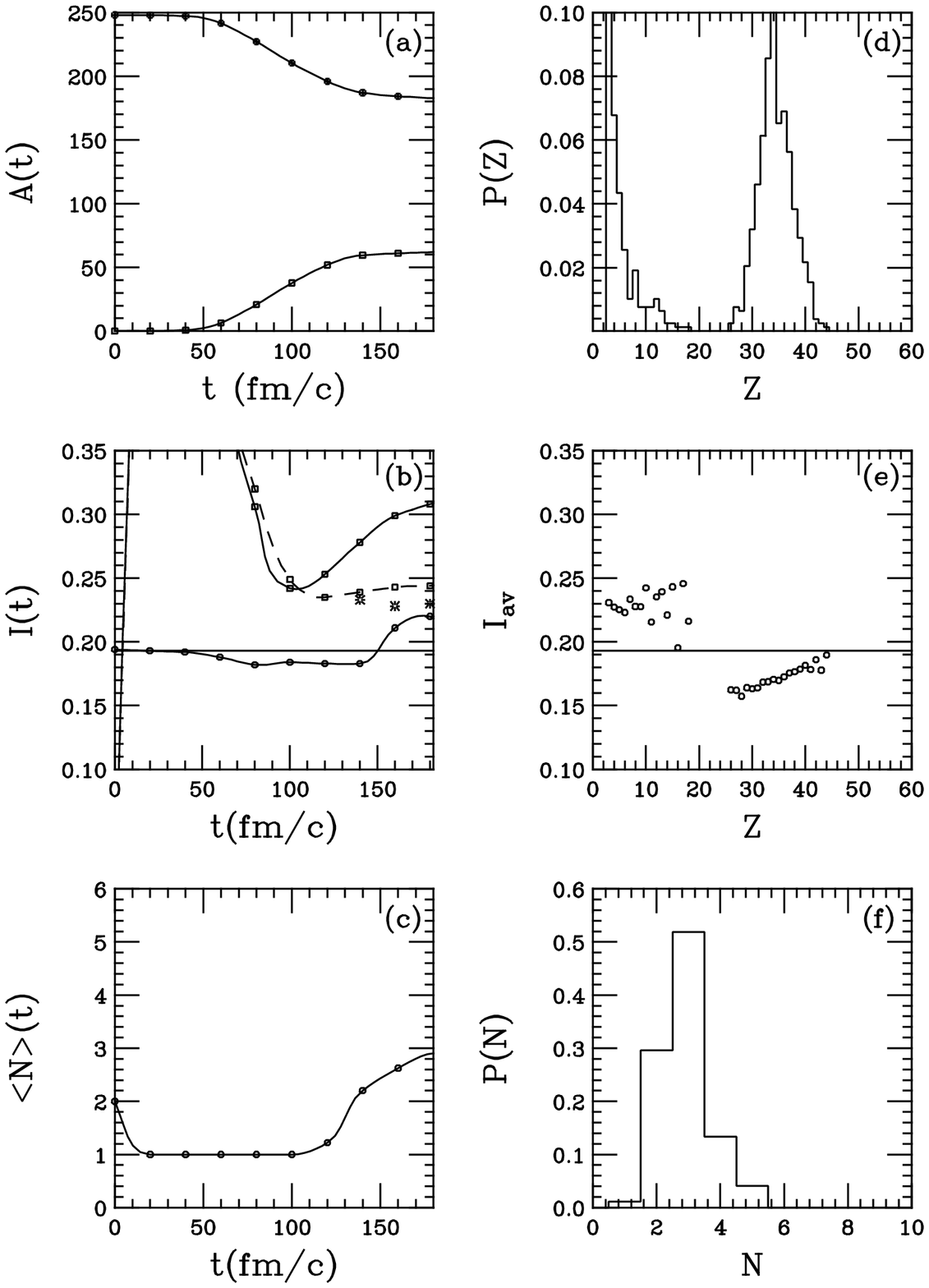}}
\caption{\it $^{124}Sn+^{124}Sn$ $b=6fm$ collision: time
evolution (left) and freeze-out properties (right).
See text. ASY-SUPERSTIFF EOS.}
\end{minipage}
\end{figure}

The differences between the results 
observed using different symmetry terms can also be 
interpreted in terms of
the corresponding density behaviours of the $n,p$-chemical
potentials, which were shown in fig. 3.
As already noted above, we can distinguish two qualitatively
different behavious in the subnuclear density region. At very low 
density, about below $0.08 fm^{-3}$, i.e. in the region 
where bulk fragmentation
and the isospin distillation effects take place
 both neutrons
and protons have the tendency to move from lower to higher density
regions. Since the variations of the two chemical
potentials are different (larger for protons) we expect a lower asymmetry in
the liquid phase \cite{nucleation} as discussed before. 
For neck fragmentation for a certain time
interval we have  contact between a more dilute phase (neck
region) and the normal density regions of the PLT/TLF. 
This happens  at density regions
between 0.08 $fm^{-3}$  and 0.16 $fm^{-3}$, where we see from fig.3
that neutrons have 
the tendency to move toward more dilute regions producing a $n$-enrichment
of the neck, while protons will migrate towards the higher density 
regions of the PLF/TLF. This explains the large difference 
in asymmetry between the
PLF/TLF and the neck fragments. 
Moreover
the neck IMF's will be always more $n-$rich compared to the fragments
produced in
the case of bulk fragmentation. The relative motion of the spectators 
as well as
surface instabilities and Coulomb effects will impose
some limitations to this interpretation, but it seems to 
give the main features and a consistent picture of the results obtained in
simulations.

It should be noticed that this effect is in addition to the geometrical 
neutron enrichment just coming from the way the neck region originates.
In fact the overlap of the surface of PLF and TLF naturally yields a
neutron-rich region, due to the neutron skin.   This effect should be
larger in the stiff parameterizations where, due to the steep behavior 
around normal density, a larger neutron skin is predicted.

Above we have discussed the relation between the average $N/Z$ of the 
IMF's produced in the neck region  and the slope 
of the symmetry term, i.e. the symmetry pressure, just
below normal density. 
A similar information is contained, in a more averaged way,
in the fragment multiplicity distributions (figs. 9f,11f, and 13f).
Multi neck-IMF events are decreasing with increasing stiffness
of the symmetry term: comparing figs. 11f and 13f we note that
in the asy-superstiff case we have $\simeq 30 \%$ more binary events.
This is indeed a signature of the larger repulsion in the
neck zone due to the larger symmetry pressure. Thus
even a  very inclusive observable, namely
fragment multiplicity, appears to be sensitive to the symmetry term
for semi-peripheral collisions.
We note that the slope of the symmetry energy around normal 
density is exactly the
symmetry pressure we expect to see active in the neutron rich
surface of heavy elements, which is thus of large importance
in determining the difference between matter and proton distributions
(see the recent discussion in refs. \cite{bro00,hor01}).
Of course, the question is absolutely critical for the discussion of
unstable nuclei and neutron skins and halos. We thus see the
complementarity between reaction and structure studies.

\subsection{Results of $^{112}Sn+^{112}Sn$, $n$-poor case}

The results for collisions of the neutron-poor system
$^{112}Sn+^{112}Sn$ are shown in the same format as before 
for the asy-stiff choice in figs. 14,15 and for the asy-soft 
choice in figs. 16,17.
Since in this system the initial asymmetry is on the $p$-rich 
side of the valley of stability there is a general trend of the
fragments to move towards stability, i.e. to asymmetries 
which are larger than the inital one. This is contrast 
to the $n$-rich case of $^{124}Sn$ where the general trend 
was towards fragments which are more symmetric than the initial
asymmetry. On this general trend are superimposed the 
mechanisms of isospin distillation and migration in a similar
way as in the $n$-rich case. Thus asymmetry distributions 
roughly appear just shifted with respect to the $n$-rich 
case (cf. e.g. figs. 14b,15b with figs. 8b,9b).  
In particular, we notice again a larger
asymmetry of the IMF's emitted from the neck in the asy-stiff case 
owing to a more efficient
simultaneous migration of protons
towards the denser regions of the PLF/TLF and of neutrons 
in the opposite direction.

\begin{figure}
\begin{minipage}[t]{67mm}
\epsfysize=9.75cm
\vspace{5cm}
\centerline{\epsfbox{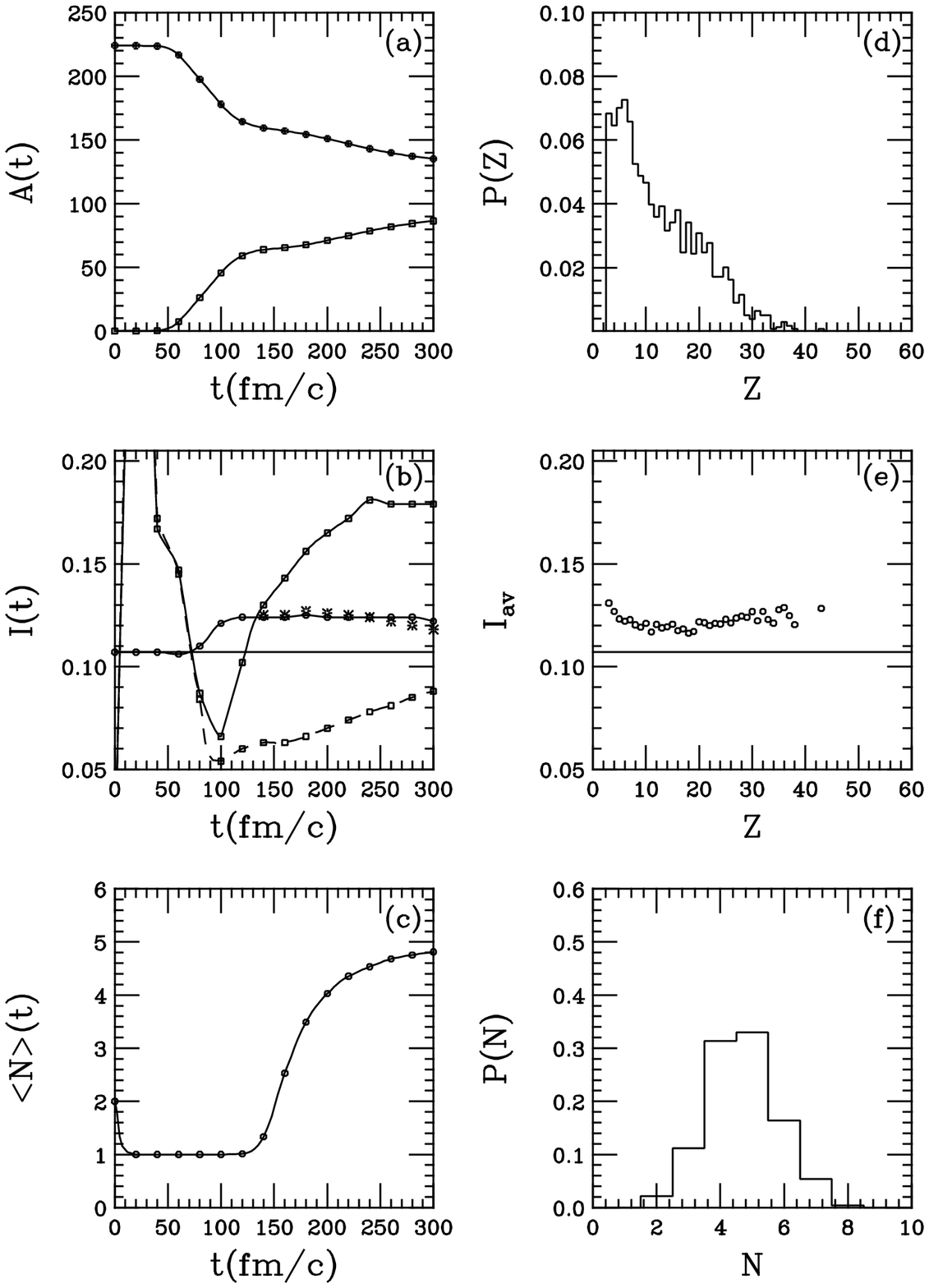}}
\caption{\it $^{112}Sn+^{112}Sn$ $b=2fm$ collision: time
evolution (left) and freeze-out properties (right).
See text. ASY-STIFF EOS}
\end{minipage}
\hspace{\fill}
\begin{minipage}[t]{67mm}
\epsfysize=9.75cm
\vspace{5cm}
\centerline{\epsfbox{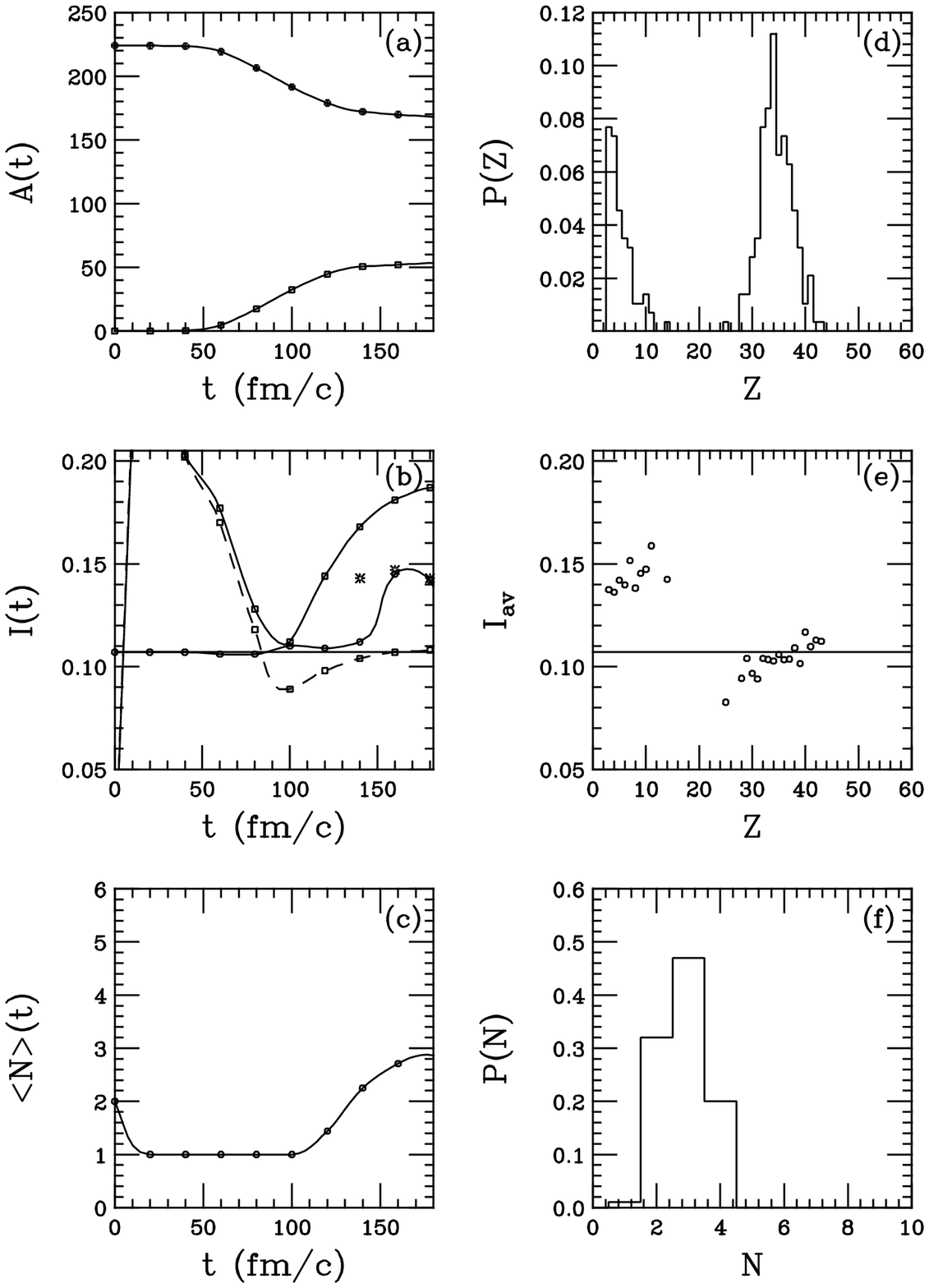}}
\caption{\it $^{112}Sn+^{112}Sn$ $b=6fm$ collision: time
evolution (left) and freeze-out properties (right).
See text. ASY-STIFF EOS.}
\end{minipage}
\end{figure}

A qualitative difference relative to the $n$-rich system is
a larger prompt proton emission during the expansion phase,
in particular for semi-central collisions as we see from the time evolution of
gas and liquid asymmetry (figs.14b,16b).
Apart from the Coulomb repulsion
the protons are less bound at subnuclear densities (fig.2).
This effect is larger for a stiffer symmetry term, as seen by
comparing the gas
asymmetry evolution in figs. 14b and 16b. Thus when the liquid phase is 
breaking up it has a larger asymmetry than the initial one and we see
an isospin distillation effect, which is almost absent in the
asy-soft case (fig.16e).

\begin{figure}
\begin{minipage}[t]{67mm}
\epsfysize=9.75cm
\vspace{5cm}
\centerline{\epsfbox{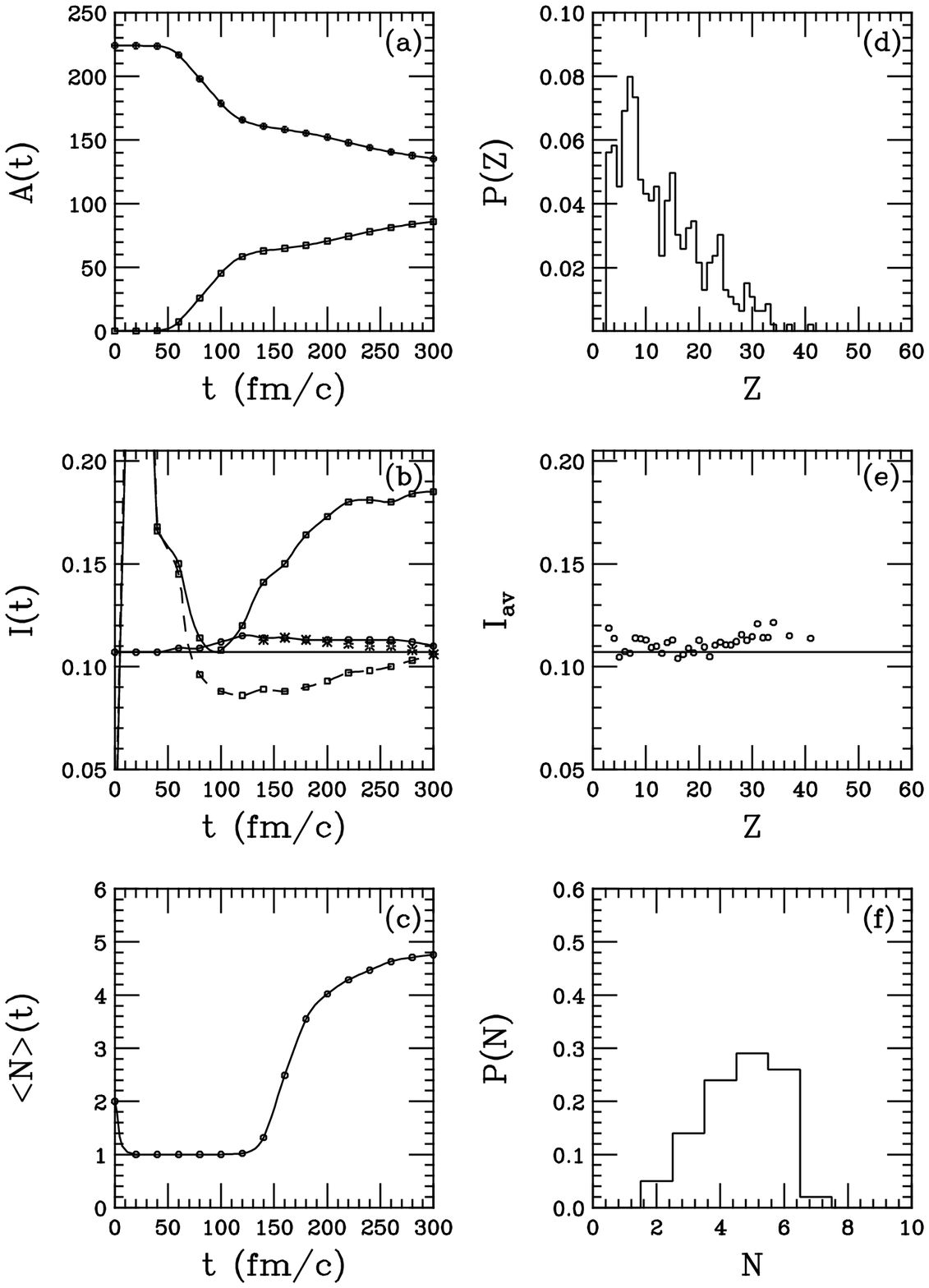}}
\caption{\it $^{112}Sn+^{112}Sn$ $b=2fm$ collision: time
evolution (left) and freeze-out properties (right).
See text. ASY-SOFT EOS}
\end{minipage}
\hspace{\fill}
\begin{minipage}[t]{67mm}
\epsfysize=9.75cm
\vspace{5cm}
\centerline{\epsfbox{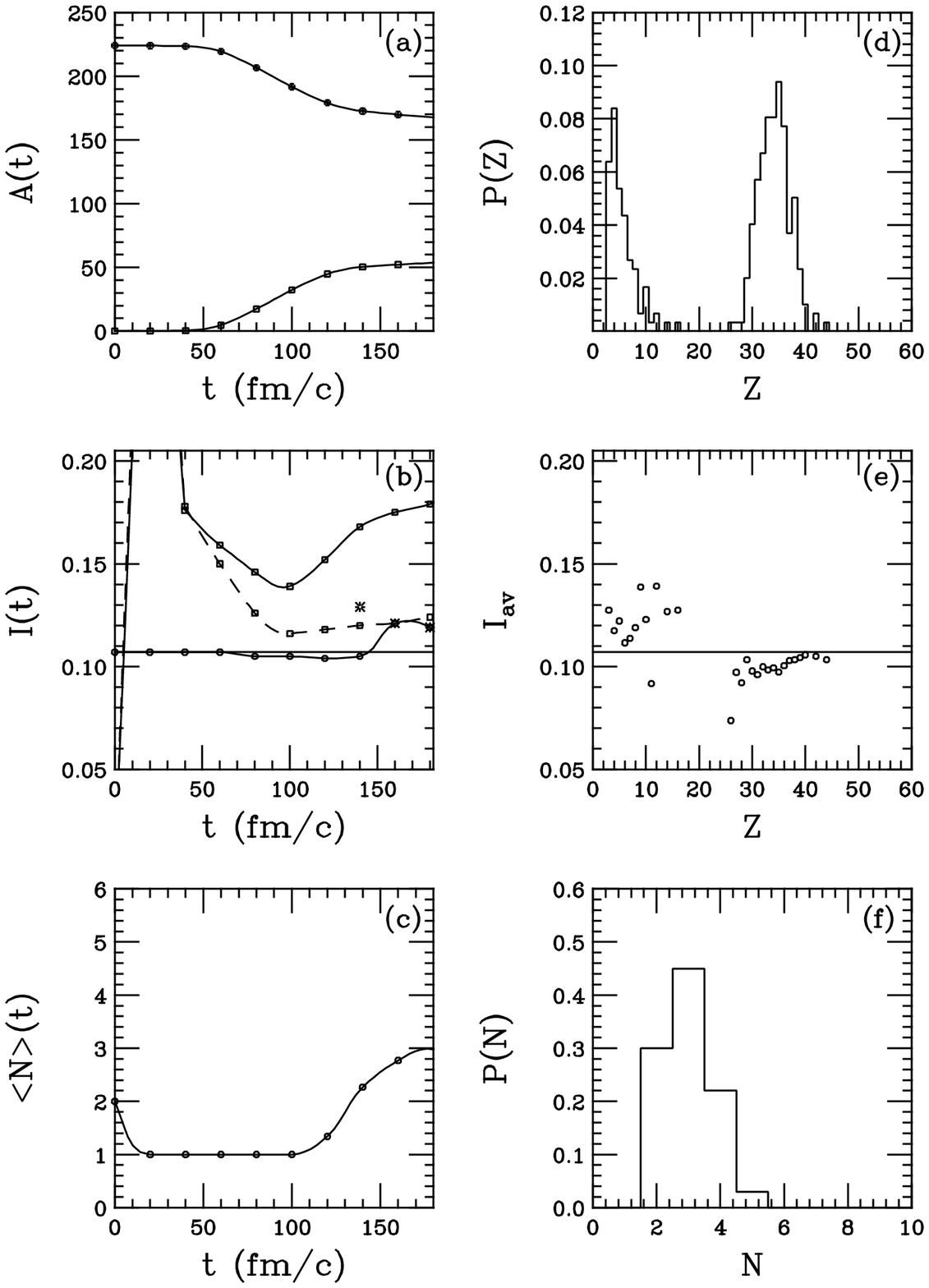}}
\caption{\it $^{112}Sn+^{112}Sn$ $b=6fm$ collision: time
evolution (left) and freeze-out properties (right).
See text. ASY-SOFT EOS.}
\end{minipage}
\end{figure}
 
In the asy-soft case
the neck isospin enrichment is produced mainly by the neutron migration
from the PLF/TLF  while the proton drift is very
reduced between 0.08 and 0.15 $fm^{-1}$ densities (see
lower dashed curve of fig.3). Therefore more protons remain
 in the low density 
region, the fragments are less asymmetric, and the difference
relative
to the PLF/TLF asymmetry is reduced (compare figs.15e and 17e).

We observe an interesting effect of the symmetry term on the 
IMF multiplicity in the bulk fragmentation.
In fig.18 we show the average
IMF multiplicity vs. the average charge of the heaviest produced fragment,
which is a measure of the centrality of the collision, for two symmetry
terms and for the two systems, $n$-rich and $n$-poor. A difference
between the multiplicities for the two systems is evident
for the asy-stiff case in semi-central collisions
(low $<Z>_{heavy}$). For the asy-soft case the difference is very
reduced. We ascribe the effect to the fact that
with a stiffer symmetry term protons in a n-rich system are very efficiently
used to form clusters, see the chemical potential argument.
Such differences have actually been observed in recent experimental data
\cite{mil99}, which could be an indication for a 
symmetry term of asy-stiff type.

\begin{figure}[htp]
\epsfysize=6.cm
\vspace{5cm}
\centerline{\epsfbox{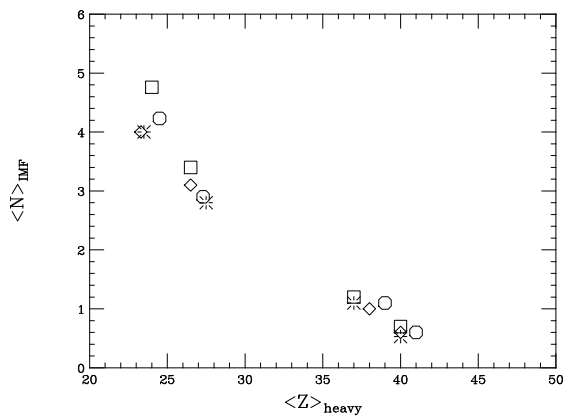}}
\caption{\it Correlation between mean IMF multiplicity and charge
of the heaviest fragment: (squares) n-rich "asy-stiff";
(diamonds) n-poor "asy-stiff"; (circles) n-rich "asy-soft";
(stars) n-poor "asy-soft".}
\end{figure}

We note that more
definitive conclusions can be obtained from precise $N/Z$ measurements
of the emitted IMF's with a good selection of the sources.
A larger absolute value of neck-IMF asymmetry in comparison
with that of fragments produced in semi-central collisions as well as a large
difference relative to that of the PLF/TLF source will strenghten
the above indication.

\section{Outlook}

In this paper we have tried to make a connection between 
chemical instabilities in an infinite binary system of
protons and neutrons and isospin transport
properties in  microscopic simulations
of fragmentation reactions of charge asymmetric ions in the
medium energy range between 20 and 100 AMeV.
We have investigated  observables that seem to be rather sensitive
to the symmetry term of the nuclear EOS. 
A first conclusion of the detailed analysis of our dynamical simulations
is that in this energy range we expect to see a characteristic
behaviour of the $N/Z$ of the intermediate mass fragments as a function 
of the centrality of the reaction, a typical {\it rise~and~fall}
with decreasing impact parameter. For peripheral collisions
IMF will be emitted in a statistical way from the excited
PLF/TLF regions close to the stability line. For semi-peripheral
events the neck-fragmentation mechanism will form  more neutron-rich 
fragments from dynamical effects. For central collisions
the neutron distillation will take place and fragments
will be again much more symmetric.
The relevance of such behaviour is related to the stiffness of the
symmetry term at subnuclear densities, and this could be a very
important information to extract from fragmentation data,
provided a very accurate centrality selection of the events is
performed.

The proton fraction of fast particle emission appears to be
quite sensitive to the slope of the symmetry term around
normal density. For central collisions this effect introduces
the observed difference in IMF's multiplicities between
$n$-rich and $n$-poor systems.

Isospin migration in the "neck instabilities" will also
induce interesting isotopic effects on the expected
fast-fission breaking of the projectile-like  and/or
target-like fragments.

Of course all the results discussed here refer to the primary
fragments, i.e. at the {\it freeze-out} time. Of course these are well
excited and the subsequent statistical decay can modify
the signal. Some of the effects discussed here appear to be 
quite robust
and indeed from the first available data the isospin dependence
of fragment production emerges quite clearly. However, a 
detailed investigation of the secondary decay, which depends 
crucially on the isospin content and the excitation energy
of the primary fragments is clearly neccessary, in order to 
make connections to experimental observations.

Moreover we have not presented here more detailed dynamical
properties of the fragment emission, like velocity distributions
and correlations,
kinetic and excitation energies, angular distributions, collective
flows, etc. The present results are already very promising and we
can be confident that with more exclusive data and more
asymmetric (also radioactive) beams we will be able to
perform detailed studies of the "elusive" symmetry term
of the nuclear EOS.

\subsection*{Acnowledgements}
We gratefully acknowledge intense and stimulating discussions with
several people actively working on isospin effects in nuclear dynamics.
In particular we would like to mention Bao-An Li, I.Bombaci, Ph.Chomaz,
 I.Hamamoto, H.Horiuchi, U.Lombardo, A.Ono, U.Schroeder, L.Sobotka
and S.Yennello. Special thanks are due to the experimental groups
working at the NSCL-MSU (W.G.Lynch, M.B.Tsang, G.Verde and R. de Souza)
and at the CS-LNS (A.Pagano, E.Geraci, E.Piaceski and J.Wilczinsky)
for the prompt availability of their inspiring new $4\pi$ data.


\begin{thebibliography}{00}
\bibitem{nucleation} We like to note that the argument is valid not
only inside the spinodal region where the nuclear phase is unstable 
but also in metastability regions were the liquid phase is developing from 
an initial homogeneous gas via "nucleation" due to large 
amplitude fluctuations.
So we expect isospin distillation also in the much 
slower nucleation mechanism.
\end{thebibliography}
\end{document}